\documentclass[usenatbib]{mn2e}   
\pdfoutput=1 
\usepackage[pdftex]{graphicx} 
\usepackage{times}
\usepackage{rotate} 
\usepackage{amssymb} 
\usepackage{rotating}
\usepackage{epstopdf} 

\title[Young radio sources in 2MASS QSOs]
{The radio spectra of reddened 2MASS QSOs: evidence for young radio jets}
\author[Georgakakis et al. ] {A. Georgakakis$^{2,1}$\thanks{email:
    age@mpe.mpg.de}, M. Grossi$^{3,4}$, J. Afonso$^{3,4}$, A. M. Hopkins$^{5}$\\
  $^1$National Observatory of Athens, I. Metaxa \& V. Paulou,
  Athens 15236, Greece\\ 
  $^2$ Max-Planck-Institut f\"ur extraterrestrische Physik, Giessenbachstrasse 1, D-85748, Garching bei M\"unchen, Germany\\
  $^3$Observat\'{o}rio Astron\'{o}mico de Lisboa, Faculdade de Ci\^{e}ncias, Universidade de Lisboa, Tapada da Ajuda, 1349-018 Lisbon, Portugal\\
  $^4$Centro de Astronomia e Astrof\'{\i}sica da Universidade de Lisboa, Lisbon, Portugal\\
  $^5$Australian Astronomical Observatory, P.O. Box 296, Epping, NSW 1710, Australia\\
}
\begin{document}
\maketitle  

\begin{abstract} Multifrequency   radio    continuum   observations
(1.4-22\,GHz) of a  sample of reddened QSOs are  presented.  We find a
high  incidence (13/16)  of  radio spectral  properties,  such as  low
frequency turnovers, high frequency spectral breaks or steep power-law
slopes, similar  to those observed in powerful  compact steep spectrum
(CSS) and gigahertz-peaked spectrum  (GPS) sources. The radio data are
consistent with relatively young  radio jets with synchotron ages $\la
10^{6} - 10^{7}$\,yr. This calculation  is limited by the lack of high
resolution (milli-arcsec)  radio observations.  For the one  source in
the sample  that such data are  available a much younger  radio age is
determined,  $\la 2\times  10^{3}$\,yr,  similar to  those of  GPS/CSS
sources.  These findings are consistent with claims that reddened QSOs
are young systems captured at the  first stages of the growth of their
supermassive black holes.  It also suggests that expanding radio lobes
may be an important feedback mode at the early stages of the evolution
of AGN.
\end{abstract}

\section{Introduction}\label{sec_intro}  

In the last few years an  increasing body of observations points to an
intimate relation between the formation  of galaxies and the growth of
the    supermassive   black    holes   (SMBH)    at    their   centres
\citep[e.g.][]{Ferrarese2000,   Gebhardt2000}.   The  nature   of  the
interplay  between the two  components is  still not  well understood,
although  important  for  understanding  both  the  star-formation  and
accretion history of the Universe.

The  large energy output  of QSOs  relative to  the binding  energy of
their hosts \citep[e.g.][]{Silk1998}  and the strong outflows observed
in  many  of  these  systems  \citep[e.g.][]{Reichard2003},  motivated
analytical  calculations  which  proposed  AGN feedback  as  the  link
between   the   formation   of   SMBH  and   the   assembly   galaxies
\citep{Silk1998,  Fabian1999,   King2003,  King2005}.   These  studies
suggest that  the energy released by  the central engine  has a strong
impact on  the interstellar medium thereby  affecting the evolutionary
path  of the  host galaxy.   These  results are  broadly supported  by
numerical  simulations, which assume  various AGN  feedback mechanisms
and     different     conditions     under     which     SMBH     grow
\citep[e.g.][]{DiMatteo2005,  DeBuhr2010,  DeBuhr2011}.   The  general
picture  emerging from  the numerical  simulations is  that  SMBHs and
stars  form  almost  simultaneously  as  the  result  of  gas  inflows
triggered by  either major mergers  \citep[e.g.][]{Springel2005}, disk
instabilities \citep{Hopkins_Quataert2010,  Bournaud2011} or shocks in
recycled     gas     from    the     winds     of    evolved     stars
\citep{Ciotti_Ostriker2007}. The  early stages  of SMBH growth  in all
those models  take place behind dust  and gas cocoons,  which at later
times  are   blown  away  by   some  form  of  AGN   related  feedback
mechanism. This  allows the central  engine to shine unobscured  for a
short period  of time,  before it  runs out of  fuel and  switches off
itself.

This   generic    evolutionary   scheme,   first    put   forward   by
\cite{Sanders1988},  has been  shown  to be  in  broad agreement  with
observations  of   AGN  and  galaxies  \citep[e.g.][]{Hopkins2008_sam,
Degraf2010, Degraf2011}.   Nevertheless many of the  details remain to
be fully understood.  A number of studies for example, debate the role
and   relative   importance   of   different   feedback   modes,   e.g
photoionisation,   radiation   pressure,   bubbles,  winds   or   jets
\citep[e.g][and           references           therein]{Hambrick2011}.
\cite{Hopkins_Elvis2010} propose a two stage feedback scheme, in which
AGN first  drive outflows  in the hot/warm  ISM, which in  turn affect
cold gas  clouds by deforming them and  hence substantially increasing
their  cross section  to ionisation  and radiation  pressure  from the
central engine. The overall effect  of this scenario is a reduction by
almost  1\,dex in  the fraction  of the  AGN energy  that needs  to be
deposited to the ISM to expel  the gas of the host galaxy and regulate
the formation of stars.  Recent studies also highlight the efficiency
of  starbursts for  depleting  the nuclear  gas  reservoirs of  galaxy
mergers,  thereby  substantially  relaxing  the requirements  for  AGN
feedback   for    quenching   star-formation   \citep[e.g.][]{Cen2011,
Hopkins2011}.

AGN  captured at  an early  stage  of their  evolution, when  feedback
processes  are expected  to  be  close to  their  peak, are  excellent
laboratories for  testing these ideas and  advancing our understanding
on  the interplay  between  the  formation of  SMBH  and their  hosts.
Reddened  QSOs, first identified  in 2MASS  \citep[Two Micron  All Sky
Survey;][]{Cutri2001,   Wilkes2002,   White2003,   Glikman2007},   are
believed to  represent young AGN.   Key properties of  this population
which are consistent with the  youth scenario include (i) red continua
most   likely    associated   with    dust   in   the    host   galaxy
\citep[e.g][]{White2003,  Glikman2007,  Urrutia2009, Georgakakis2009},
(ii)  a high  incidence  of morphological  disturbances suggestive  of
recent   mergers  \citep{Urrutia2008},  (iii)   optical  spectroscopic
signatures for  fast outflows  in the warm  phase of  the interstellar
medium   \citep{Urrutia2009},  (iv)   enhanced   star-formation  rates
relative  to  UV   bright  QSOs  \citep{Georgakakis2009}.   All  these
findings  are consistent  with  the scenario  that  reddened QSOs  are
observed shortly before  or during the blow-out of  their natal cocoon
of dust and gas, i.e. at the  stage where AGN feedback is close to its
peak.

In Georgakakis et  al. (2009) a higher radio  detection rate was found
for  reddened QSOs compared  to UV bright  ones.  First,  this suggests
that expanding  radio lobes  may play an  important role at  the early
stages of  SMBH growth.  Second, by determining  the synchrotron age
of  the electron  population  responsible  for the  jets  one can  set
additional independent constraints to  the youth scenario for reddened
QSOs. The  typical characteristics of young radio  sources are compact
radio sizes  (less than few tens  of kpc), radio spectra  which show a
turnover at low frequencies (attributed to synchrotron self-absorption
or free-free  absorption) and/or a double  power-law distribution with
relatively  flat power-law  slope at  centimetre  wavelengths ($\alpha
\approx  0.7$,   where  the  flux   density  at  frequency   $\nu$  is
$S_{\nu}\propto\nu^{-\alpha}$)  followed  by  a steepening  at  higher
frequencies \citep{ODea1998}.   In the standard  theory of synchrotron
emission from expanding  lobes \citep{Kardashev1962}, spectral indices
steeper than  the canonical value  of 0.7 are attributed  to spectral
aging.  In this picture the break frequency, at which the slope of the
power-law spectrum steepens, decreases with the time elapsed since the
formation of the source as $\nu_{br}\propto t^{-2}$.  The shape of the
radio spectrum can therefore constrain the age of the jet.

This paper  presents VLA  (Very Large Array)  and EVLA  (Expanded Very
Large Array) continuum observations in the frequency range 1.4-22\,GHz
of reddened  QSOs selected  from the literature.   This multifrequency
dataset  is used  to  (i) constrain  the  overall shape  of the  radio
spectra of  reddened QSOs to  search for features indicative  of young
radio jets  (e.g. spectral breaks,  low frequency turnovers)  and (ii)
estimate the break  frequency  $\nu_{br}$ to  use  it  as an  age
indicator. Additionally by constraining  the radio spectra of reddened
QSOs one  can place them in  the context of the  population of compact
steep  spectrum  (CSS)   and  gigahertz-peaked  spectrum  (GPS)  radio
sources. These objects are proposed  to host young and expanding radio
lobes \citep[e.g.][]{ODea1998, Murgia1999}, which are likely to have a
strong    impact   on    the    ISM   of    their   parent    galaxies
\citep[e.g.][]{Holt2011}.  Throughout the paper  we adopt $\rm H_{0} =
70  \, km  \, s^{-1}  \, Mpc^{-1}$,  $\rm \Omega_{M}  = 0.3$  and $\rm
\Omega_{\Lambda} = 0.7$.

\section{Observations}\label{sec_observations}

The reddened  QSOs followed with  radio continuum observations  at the
VLA and EVLA  are selected from the samples  of \cite{Urrutia2009} and
\cite{Georgakakis2009}.   These  studies  combined  the 2MASS  in  the
near-infrared and the  Sloan Digital Sky Survey (SDSS)  in the optical
to  select  sources   with  red  optical/near-infrared  colours  (i.e.
$R-K>5$,   $J-K>1.3$\,mag   in   Urrutia   et  al.    2009;   $R-K>5$,
$J-K\ga2$\,mag in Georgakakis et al.  2009).  The sample of Urrutia et
al.  (2009)  also includes pre-selection  of 2MASS sources  with radio
counterparts in the FIRST survey  \citep[Faint Images of the Radio Sky
at   Twenty-Centimeters,][]{Becker1995}.   The  2MASS   reddened  QSOs
targeted  by  our  VLA  and  EVLA  programmes  are  listed  in  Tables
\ref{tab_sample_vla} and  \ref{tab_sample_evla} respectively. They are
selected to have radio  (1.4\,GHz) flux densities $S_{1.4} \ga 5$\,mJy
in the FIRST survey.

\subsection{VLA  data} Multi-frequency  (1.4, 4.85,  8.4  and 22\,GHz)
simultaneous   observations   of   the   sources   listed   in   Table
\ref{tab_sample_vla}  were carried  out at  the VLA  in  snapshot mode
between May  and August 2009.   The VLA configurations  changed during
that  period  from  CnB  to C.   Targets  2MASSQSO\,10,  F2MS0832+050,
F2MS0932+385 were observed in the CnB configuration with the remainder
observed in the C configuration. The dwell time in the L (1.4\,GHz), C
(4.85\,GHz)  and  X (8.4\,GHz)  bands  was  30\,s.   The targets  were
bracketed  by observations  in the  same frequency  of a  nearby ($\rm
<10\,deg$) phase  calibrator. In  the K-band (22\,GHz)  fast switching
phase calibration was used to minimise phase variations.  A cycle time
of 100 seconds (70\,sec on  source, 30\,sec on calibrator) and a dwell
time of 5\,min  was adopted. Observations in the  K-band were preceded
by pointing  observations in the  X-band of the  secondary calibrator.
This is  to minimise pointing  errors which can  significantly degrade
the sensitivity in the K-band as the a priori pointing can be off by a
large  fraction  of the  primary  beam  in  that band.   Primary  flux
calibrators  (3C147  or 3C286)  were  observed  in  all bands  at  the
beginning of the multifrequency  observations sequence of each target.
Source  F2MS\,1618+350 and F2MS\,1540+492  lack 1.4  GHz observations,
while F2MS\,1615+ 031 has not been observed at 22\,GHz.

The data  were reduced using the Astronomical  Image Processing System
(AIPS) maintained by NRAO and following standard reduction procedures.
The secondary flux calibrators close to the targets were used to track
gains and phases of each  antenna. The calibration solutions have been
derived with  the task {\sc calib}  and they have been  applied to the
targets  with   the  task   {\sc  clcal}.   The   K  band   data  were
self-calibrated using standard procedures to improve the image dynamic
range  and  reduce  the  noise  level.   Typically  one  iteration  of
phase-only  self-calibration  was performed.   Finally  the task  {\sc
  imagr},  which  includes the  algorithm  {\sc  clean},  was used  to
deconvolve the images.

The final  images noise levels range from 0.3  to 0.8 mJy/beam (L
band), 0.2 to 0.3 mJy/beam (C and  X bands) and 0.3 to 0.5 mJy/beam (K
band).   The Gaussian restoring  beam FWHM  (Full Width  Half Maximum)
varies for  each set of observations  at each band.   The typical beam
sizes are approximately $19\arcsec\times 15 \arcsec$, $5\arcsec \times
4  \arcsec$,  $3  \arcsec  \times  2  \arcsec$  and  $1\arcsec  \times
0\farcs8$,  at 1.4,  4.8,  8.4 and  22\,GHz  respectively. The  linear
extent of the beam size  of the highest spatial resolution radio image
(22\,GHz and if  not available 8\,GHz) at the  redshift of each source
is  listed in  Table \ref{tab_sample_vla}.   Flux densities  have been
measured by fitting  Gaussians to the source profiles,  using the AIPS
task JMFIT.  The results are presented  in Table \ref{tab_sample_vla}.
That  table also  lists  the  rest-frame 5\,GHz  radio  power of  each
reddened QSO.

For  sources  F2MS\,1540+492,   F2MS\,1618+350,  which  lack  1.4  GHz
observations in  our VLA program,  we use the 1.4\,GHz  flux densities
estimated by  the FIRST  and Glikman et  al. (2007)  respectively. For
source F2MS\,1618+350  the FIRST and  Glikman et al.   (2007) 1.4\,GHz
flux densities differ by a factor of about 2.  This discrepancy may be
because of variability, extended emission which is resolved out by the
FIRST  data or calibration  problems. Our  data at  higher frequencies
(4.8, 8.4,  22\,GHz) do not  show extended radio emission.  Glikman et
al.  (2007)  have  also   determined  the  8.4\,GHz  flux  density  of
F2MS\,1618+350,  which is in  good agreement  with our  estimate. This
suggests that any biases because of source variability between the two
observation epochs are small.  We  therefore choose to use the Glikman
et al.  (2007) 1.4\,GHz flux  density for F2MS\,1618+350. It is noted
that for the majority of our sample sources the 1.4\,GHz  densities
estimated  by  FIRST   and  our     observations typically agree
within about 30  per cent. 

Also, all  sample sources,  except F2MS\,1618+3502, are  unresolved in
the FIRST or our VLA observations, suggesting sizes smaller than a few
arcseconds.  Source F2MS\,1618+3502  shows two  jets extending  out to
about 20\,arcsec  (115\,kpc) from a  radio bright core.  Another 2MASS
QSOs  in Table  1, F2MS1030+5806,  has been  observed  at milli-arcsec
resolution  at  5\,GHz.   The  data  are from  the  VLBA  Imaging  and
Polarization  Survey (VIPS)  presented by  \cite{Helmboldt2007}.  This
source has  a very  compact radio morphology  with a small  scale (few
milli-arcsec) radio jet with linear size at the redshift of the source
of about 100\,pc.

\subsection{EVLA data}

 The EVLA data  were obtained in snapshot  mode at the L, C,  X, and K
bands  between  the  13th  and  the   30th  of  May  2010,  in  the  D
configuration.   Observations  were  carried  out in  two  contiguous,
128-MHz IF bands,  each divided into 64 channels,  and centred at 1348
and  1860\,MHz  (L-band),  4896   and  5024\,MHz  (C-band),  8396  and
8524\,MHz  (X-band), 22396 and  22524\,GHz (K-band).   Observations at
22\,GHz have been carried out  in the fast switching phase calibration
mode to minimise phase variations.

The  EVLA  data  were  reduced  using the  Common  Astronomy  Software
Applications  (CASA) package  release  3.0.2, maintained  by NRAO  and
following the standard reduction procedure.  Bandpass and flux density
calibration were performed using the primary flux calibrators close to
the targets (3C147, 3C286).  The  amplitude scale was set according to
the Perley-Butler coefficients derived from recent measurements at the
EVLA.   Amplitude and  phase  gains were  derived  for all  calibrator
sources,  with the task  GAINCAL, and  they have  been applied  to the
targets with the  task APPLYCAL, while the final  images were produced
with the  task CLEAN using a  Briggs weighting.  The  EVLA K-band data
did not need to be self-calibrated.

The final  images noise levels range in  the intervals 0.3-1\,mJy/beam
(L-band),  0.2-0.3 mJy/beam  (C-band), 0.1-0.2  mJy/beam  (X-band) and
0.06-0.2 mJy/beam (K-band). The  Gaussian restoring beam FWHM slightly
varies for  each set of observations  at each band.   The typical beam
size   is  $1\farcm2  \times   0\farcm8$,  $20\arcsec\times14\arcsec$,
$11\arcsec\times9\arcsec$, $4\arcsec \times 3\arcsec$, for the L, C, X
and  K band,  respectively.  The  linear extent  of the  beam  size at
22\,GHz (highest spatial resolution) at the redshift of each source is
listed  in  Table  \ref{tab_sample_evla}.   Flux densities  have  been
measured by fitting  Gaussians to the source profiles,  using the task
IMFIT. In the  L-band we chose to produce  two independent images, one
for each IF  unit, instead of combining the data.   This is because of
the wide  separation of  the central frequencies  of the two  IF bands
(1348   and  1860\,MHz).    The   results  are   presented  in   Table
\ref{tab_sample_evla}.  That table  also lists  the  rest-frame 5\,GHz
radio power of each reddened QSO.

Source  F2MS0841+3604  has not  been  observed  in  the X-band.   Also
technical problems affected  the L-band observations of F2MS0841+3604,
F2MS0915+2418,  F2MS1456+0114. As  a  result it  was  not possible  to
obtain reliable measurements at either 1.3\,GHz or 1.8\,GHz. For those
sources the FIRST radio flux density at 1.4\,GHz is adopted (see Table
2).

\section{Radio spectra of 2MASS QSOs}\label{results}

\begin{figure*}
\begin{center}
\includegraphics[height=0.5\columnwidth]{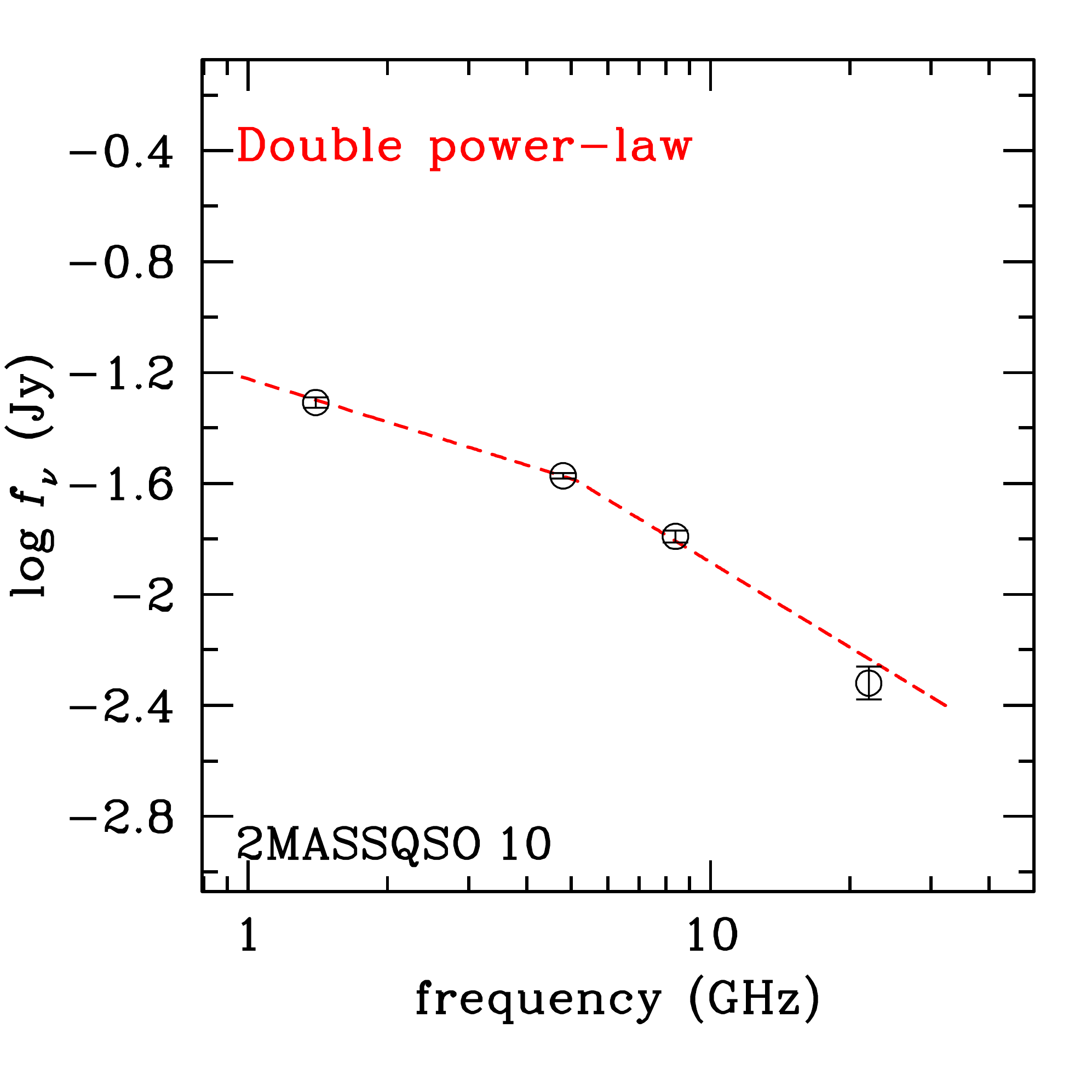}
\includegraphics[height=0.5\columnwidth]{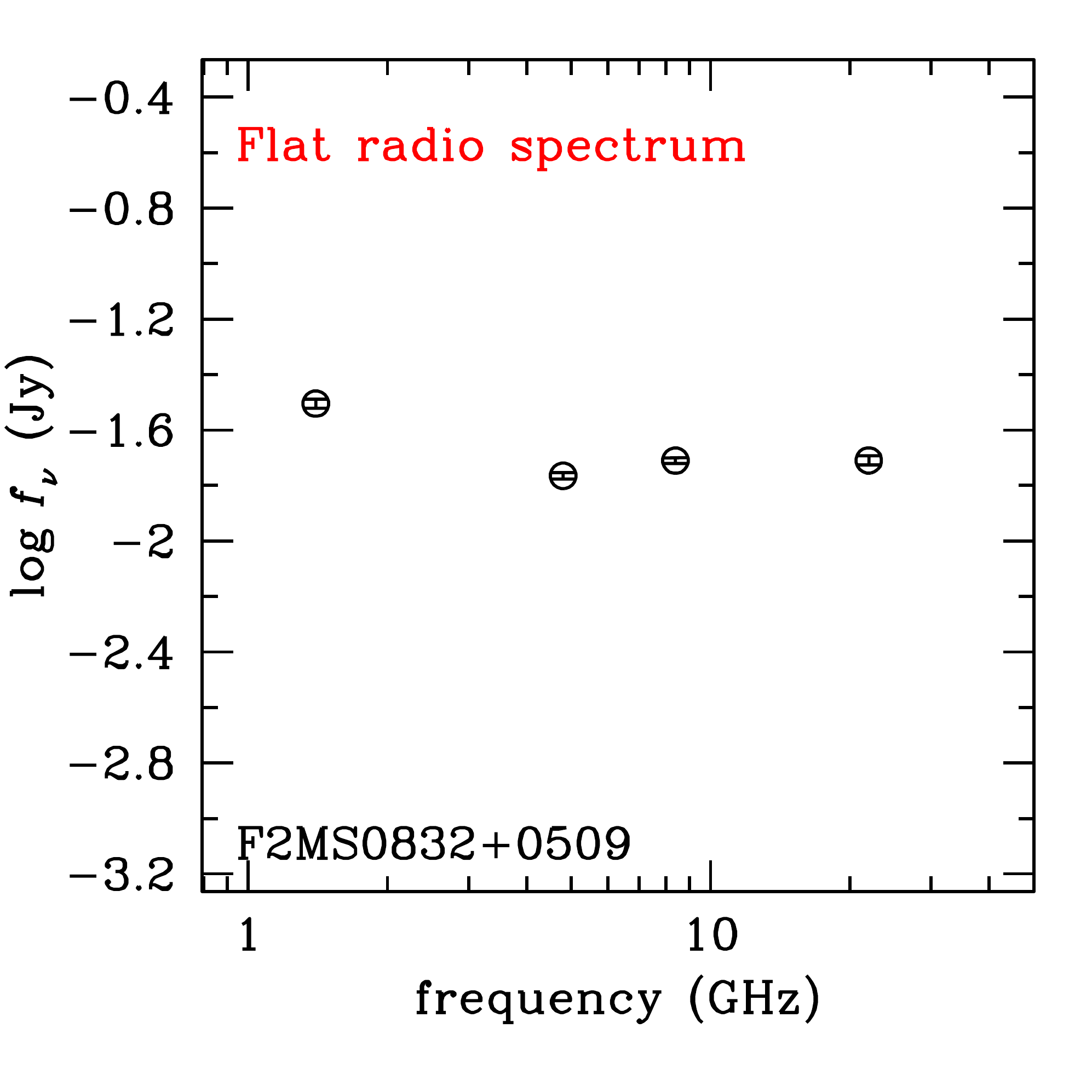}
\includegraphics[height=0.5\columnwidth]{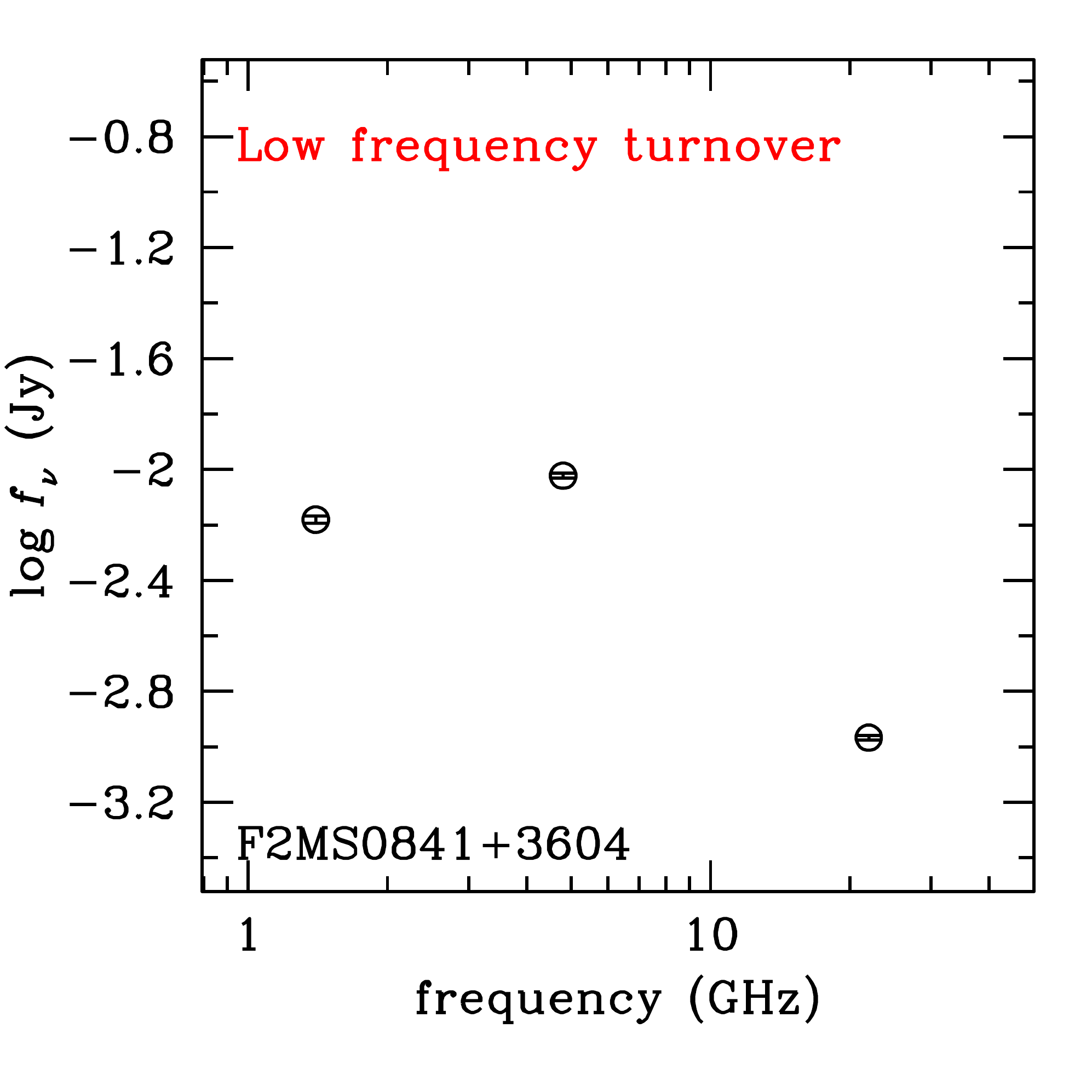}
\includegraphics[height=0.5\columnwidth]{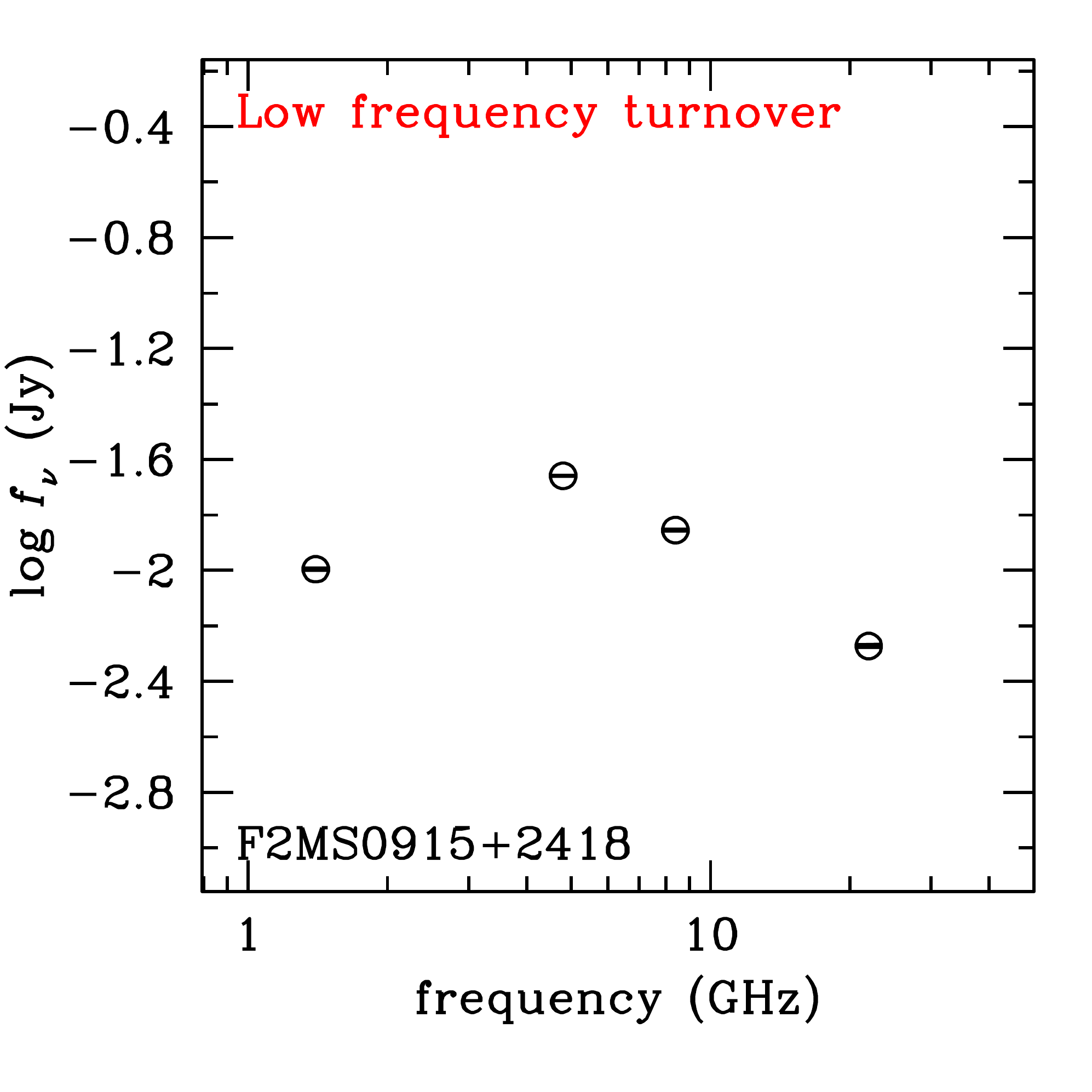}
\includegraphics[height=0.5\columnwidth]{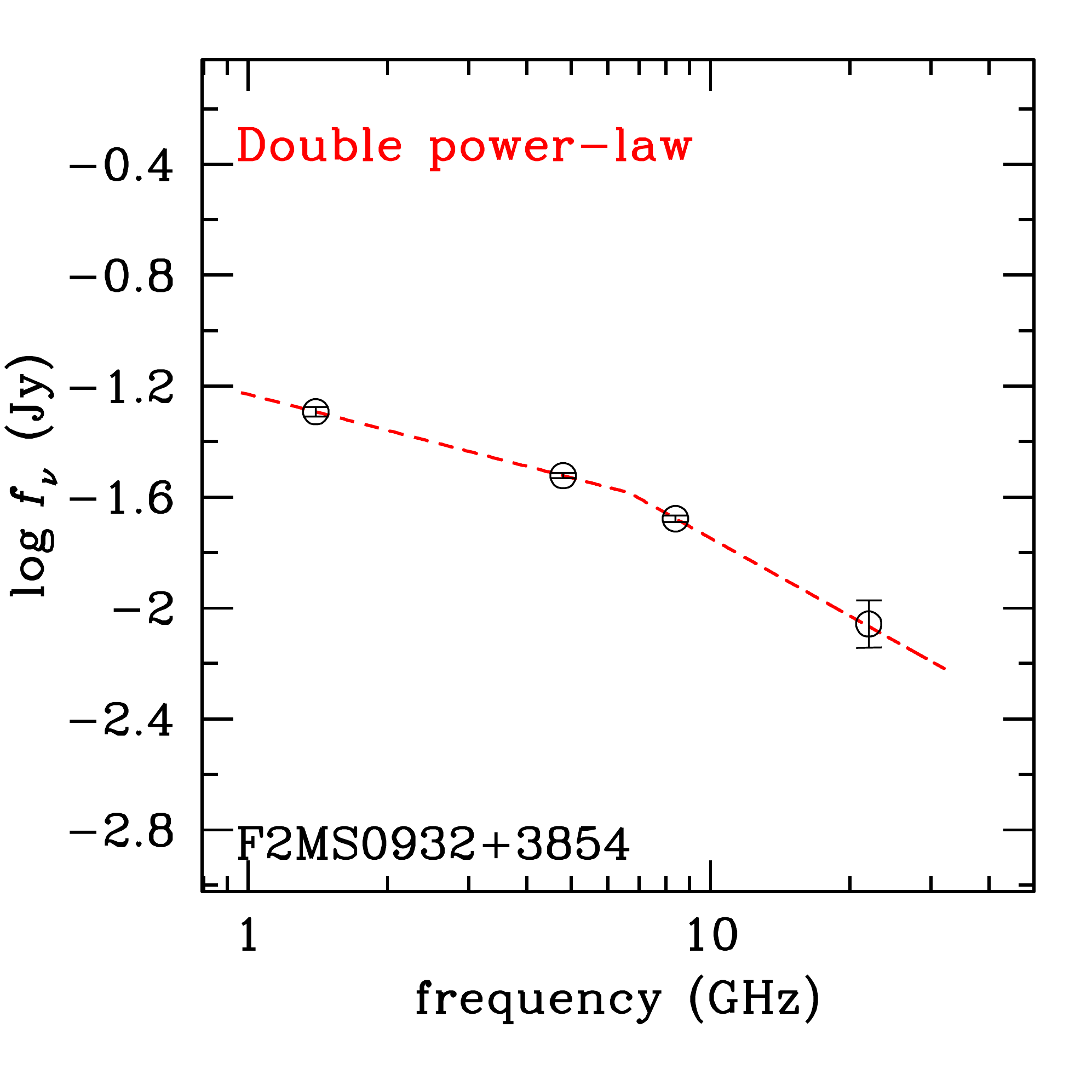}
\includegraphics[height=0.5\columnwidth]{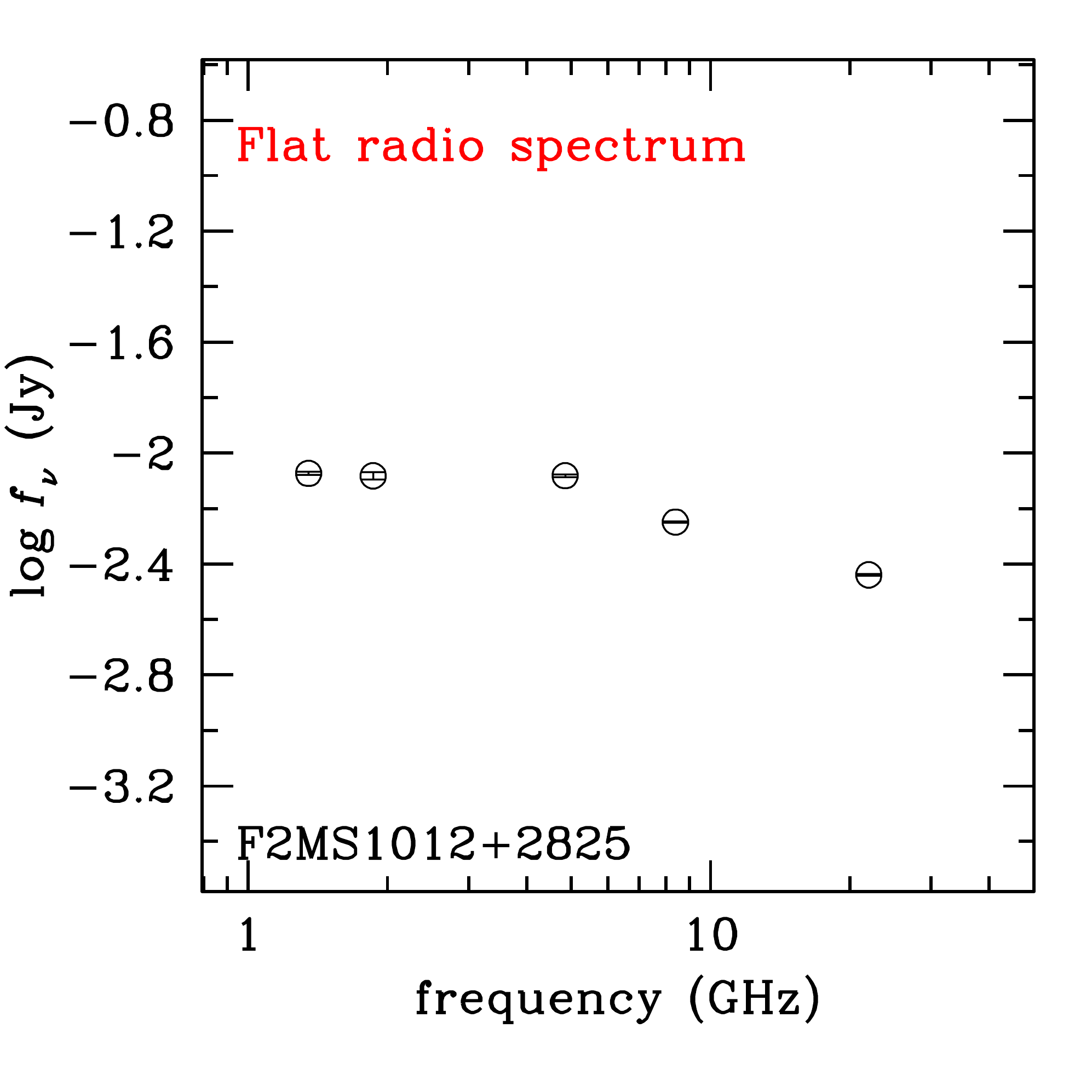}
\includegraphics[height=0.5\columnwidth]{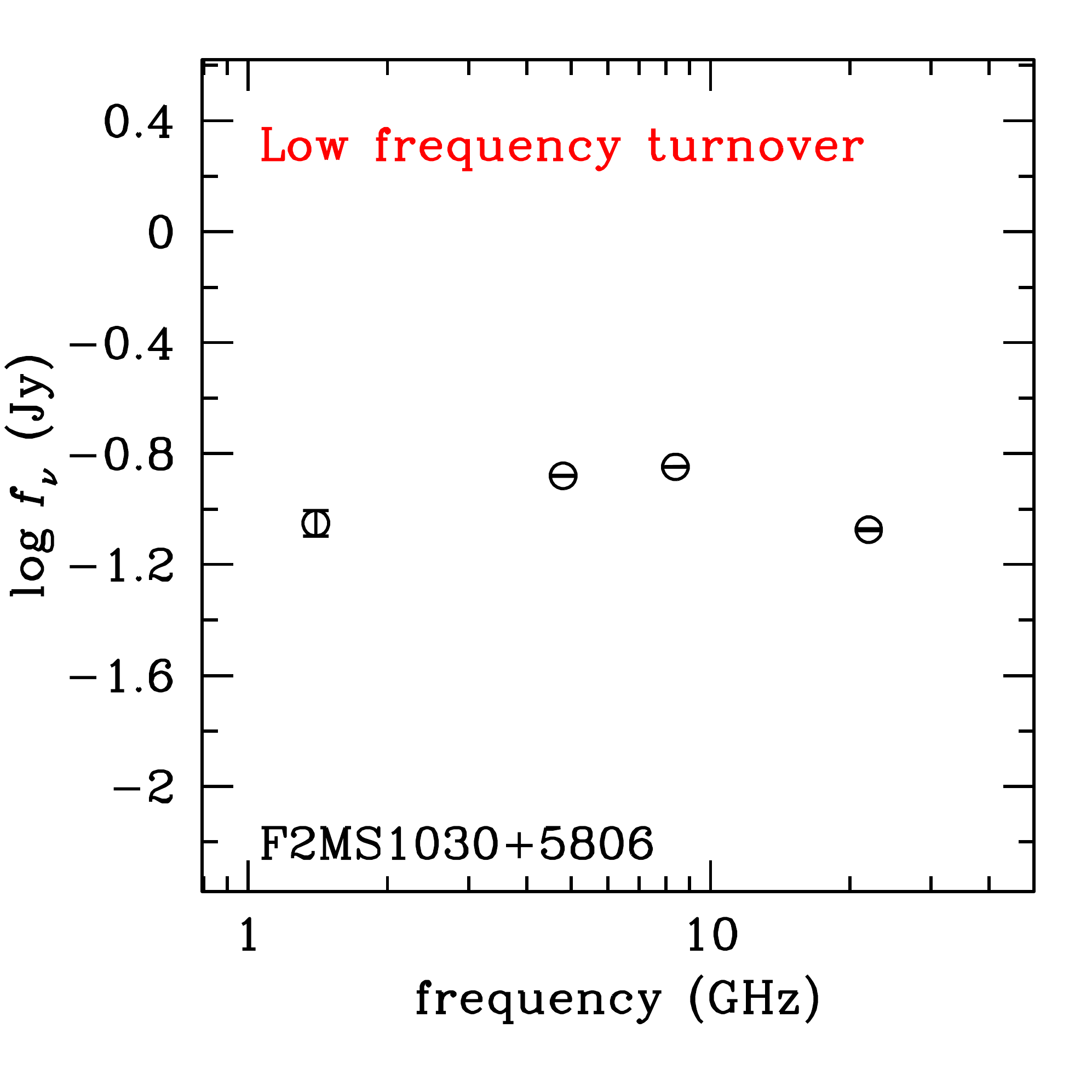}
\includegraphics[height=0.5\columnwidth]{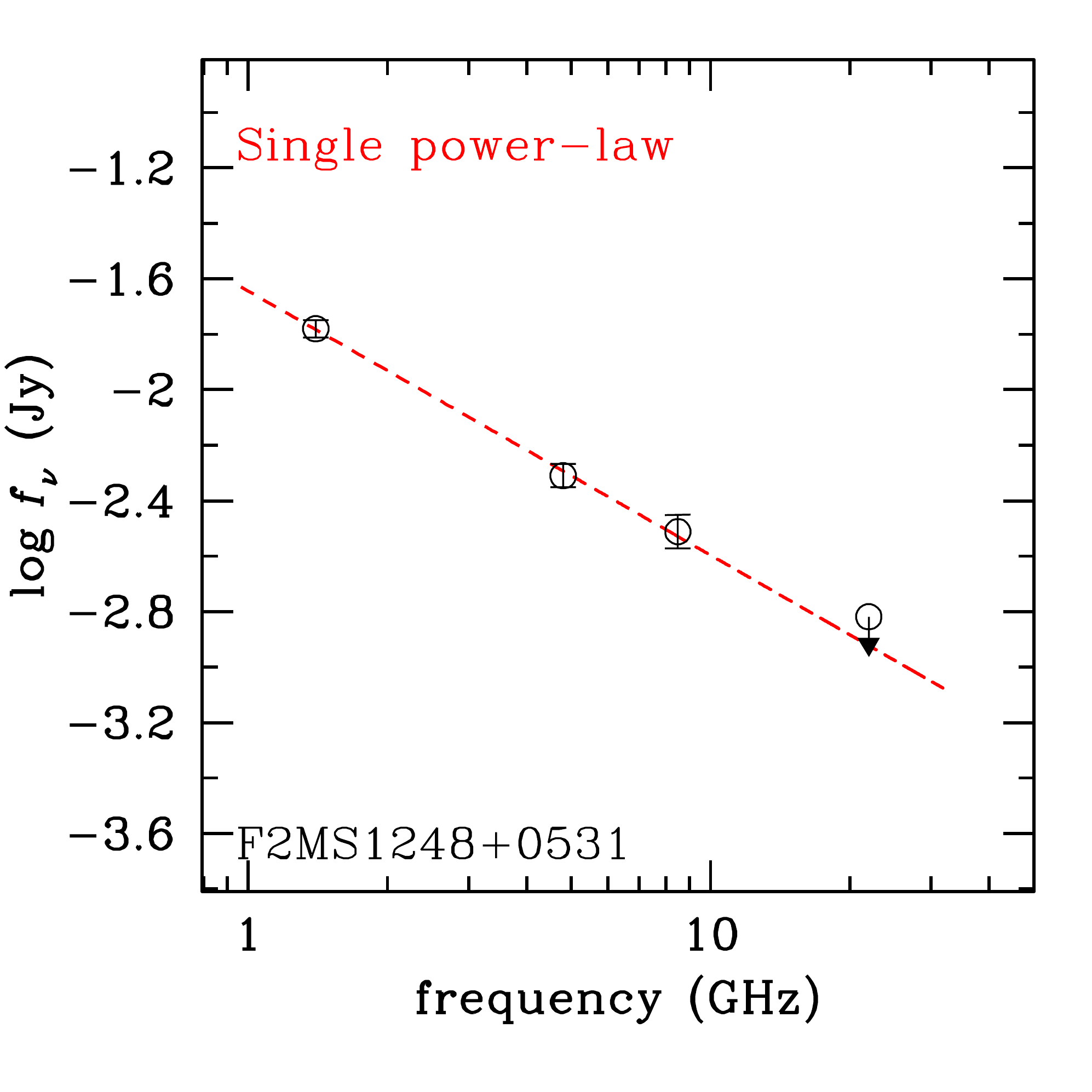}
\includegraphics[height=0.5\columnwidth]{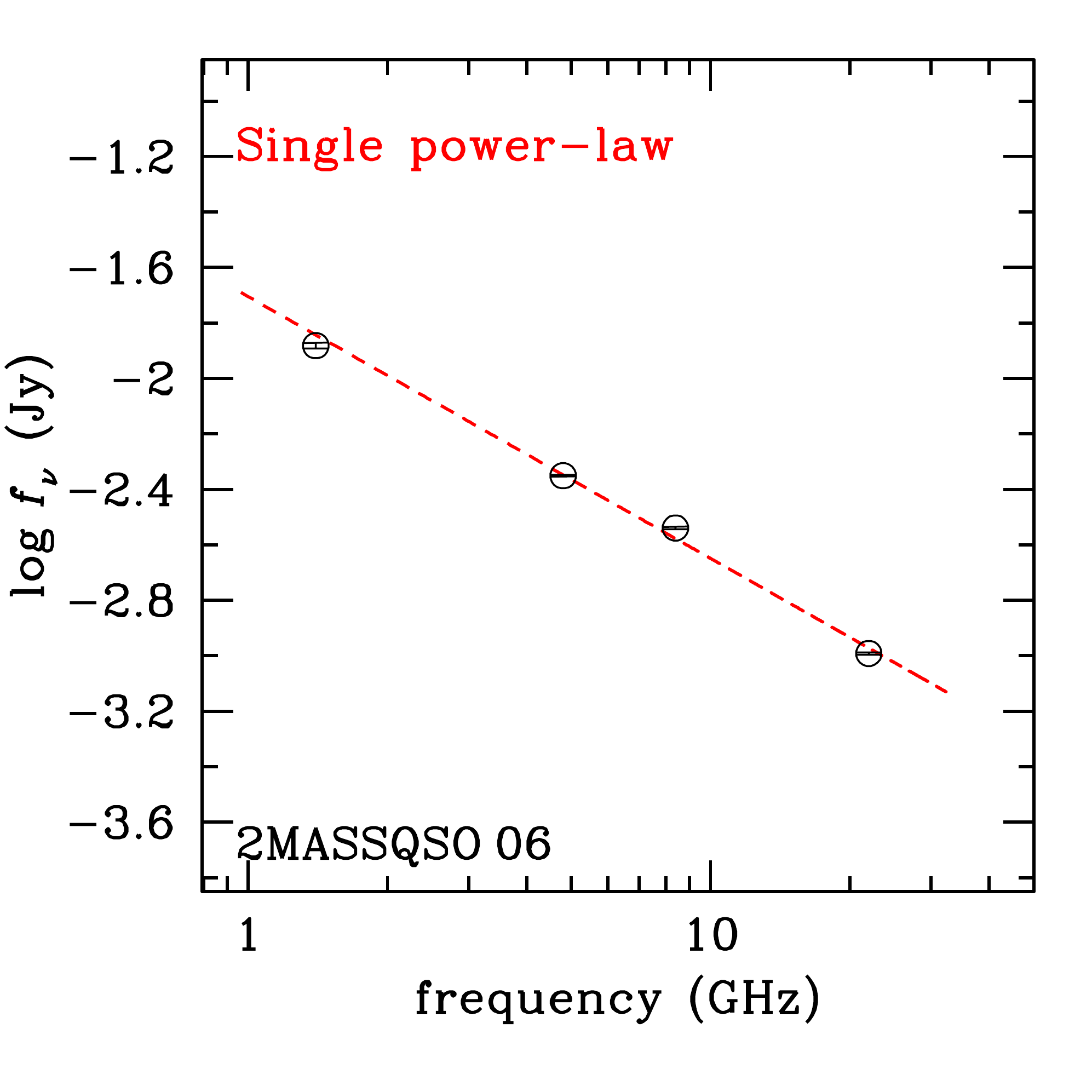}
\includegraphics[height=0.5\columnwidth]{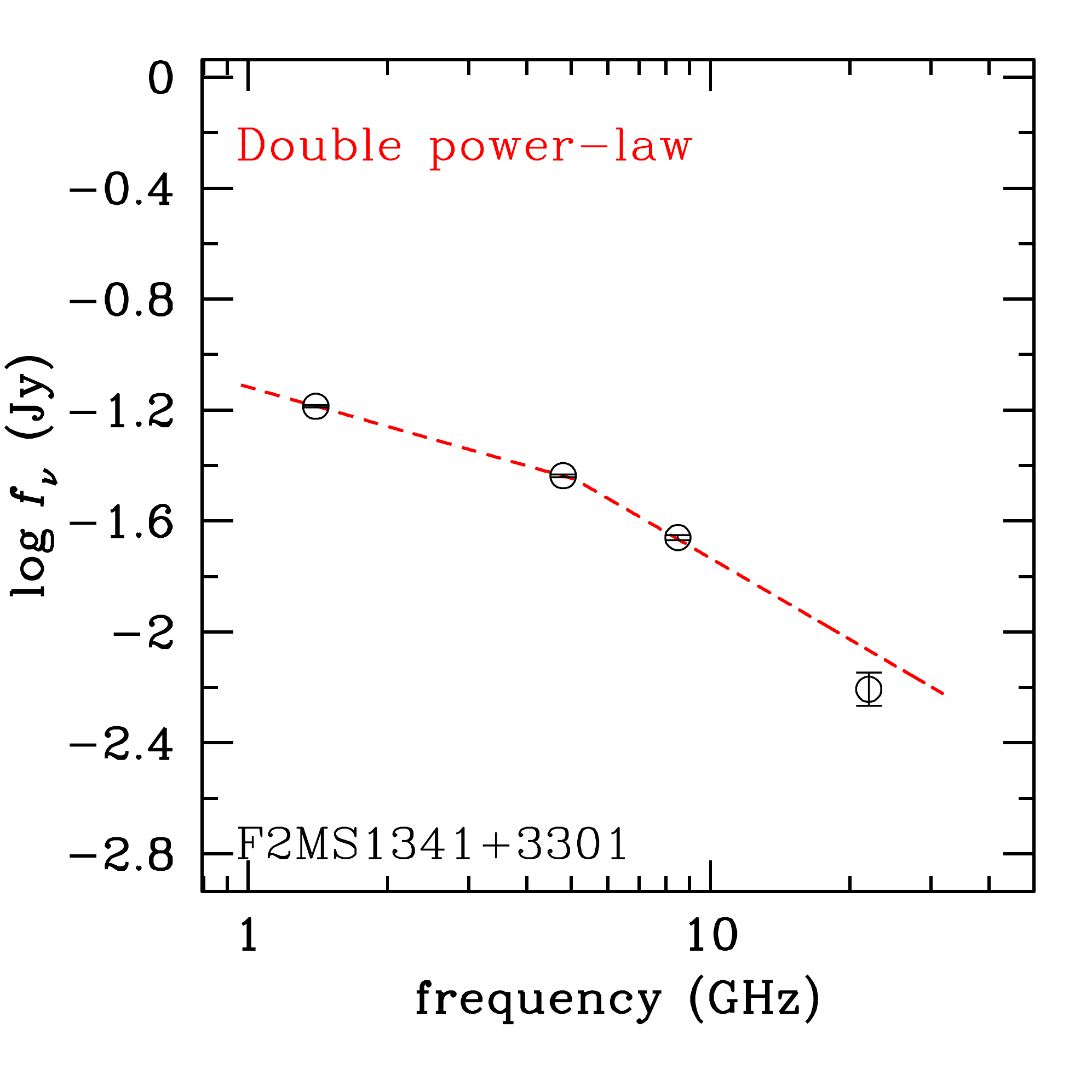}
\includegraphics[height=0.5\columnwidth]{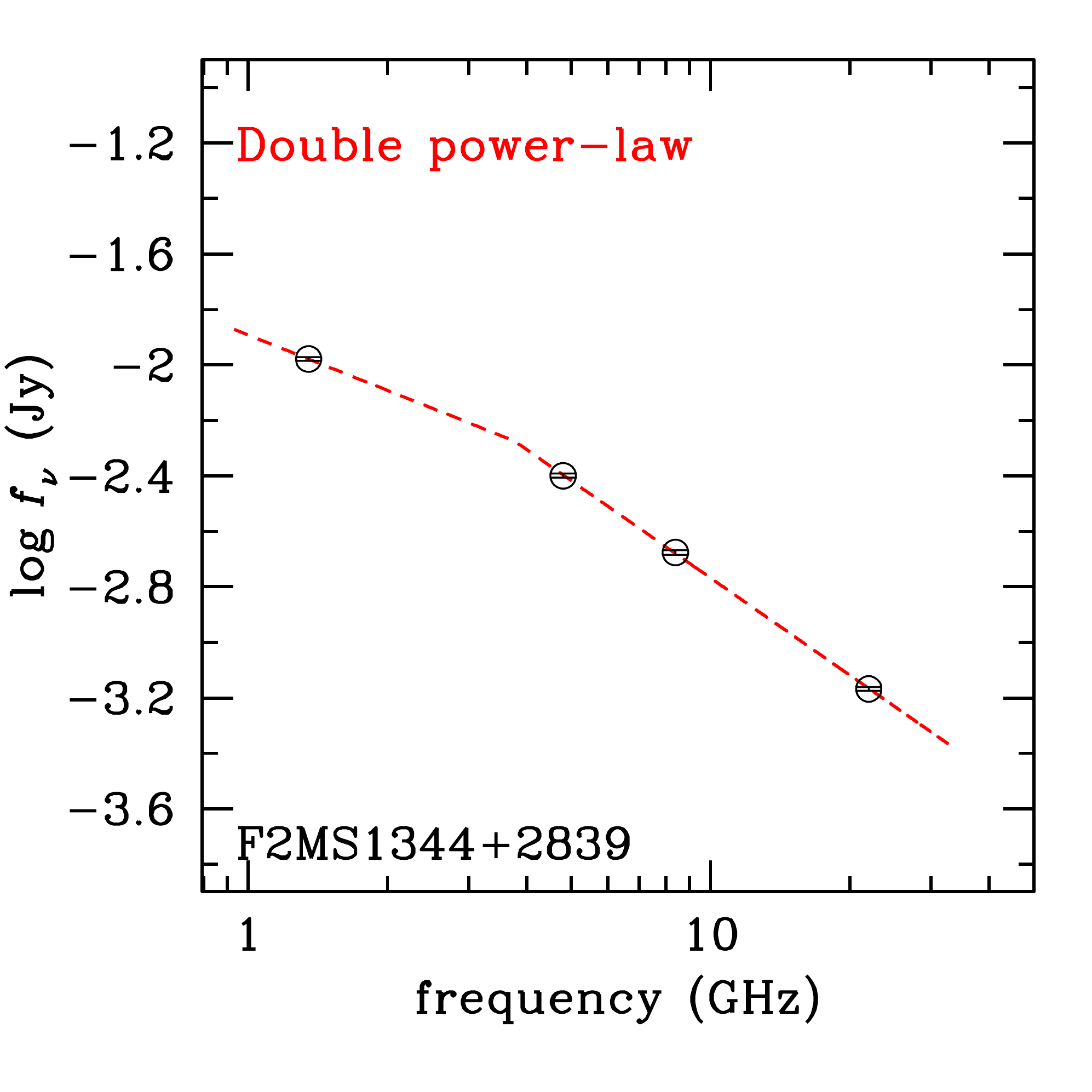}
\includegraphics[height=0.5\columnwidth]{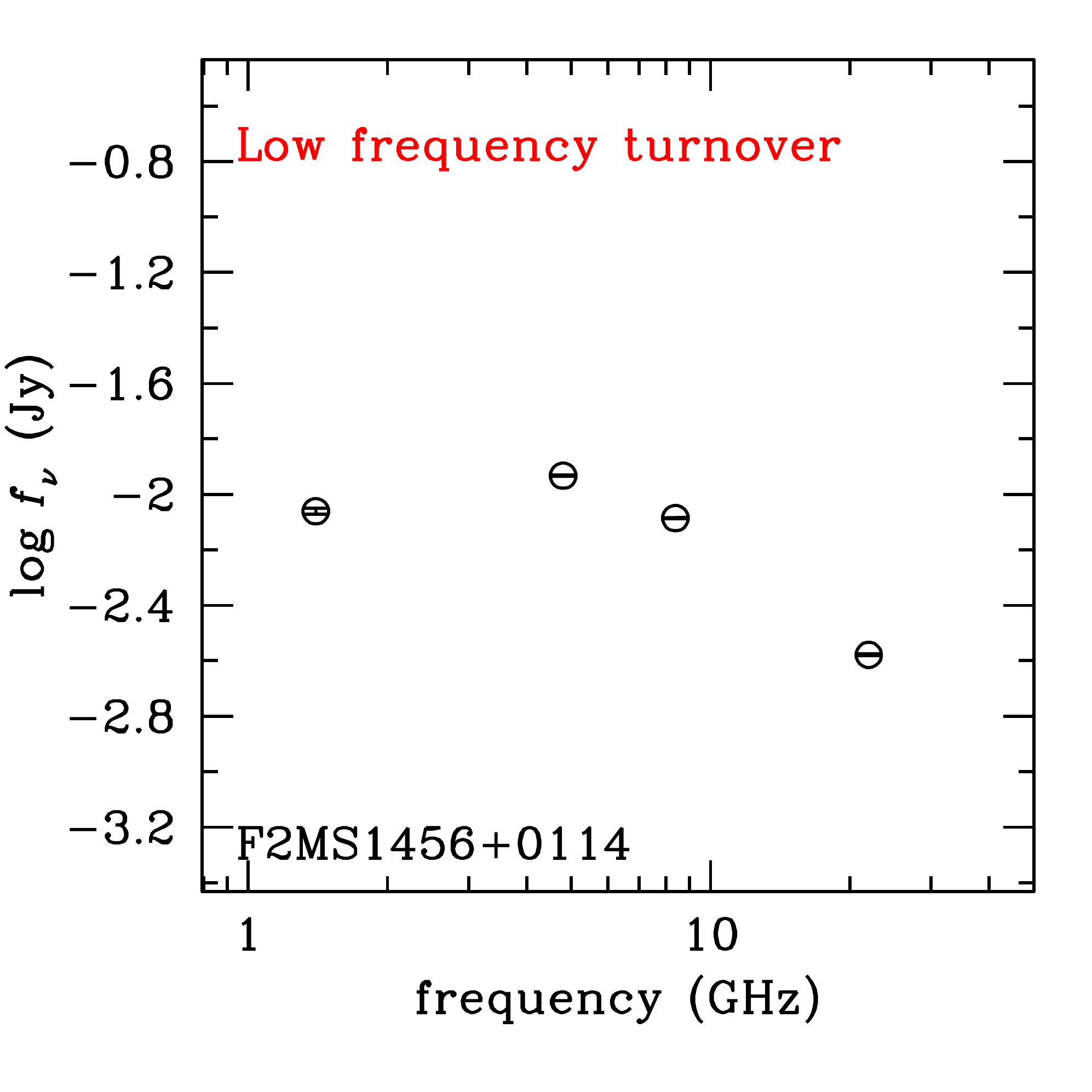}
\includegraphics[height=0.5\columnwidth]{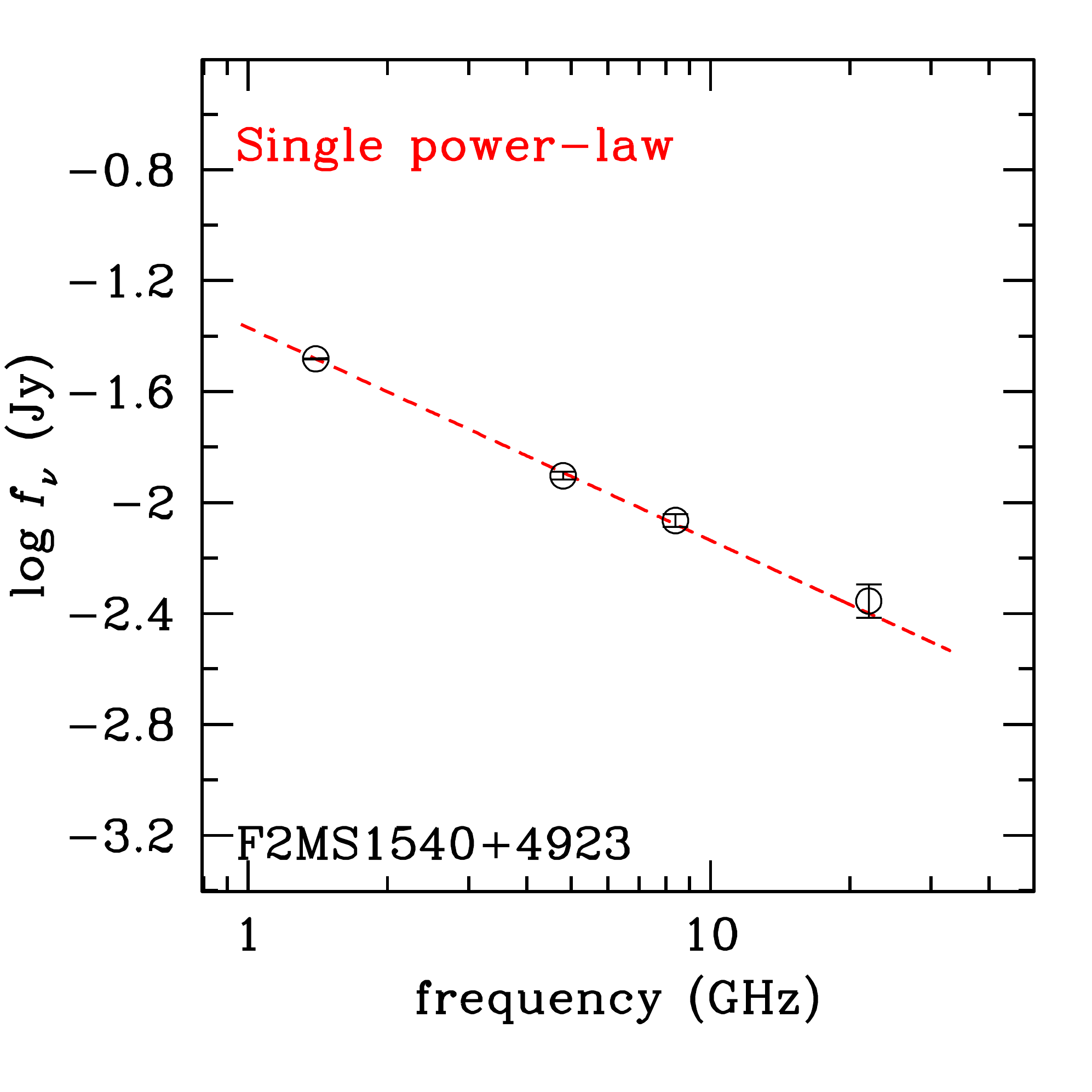}
\includegraphics[height=0.5\columnwidth]{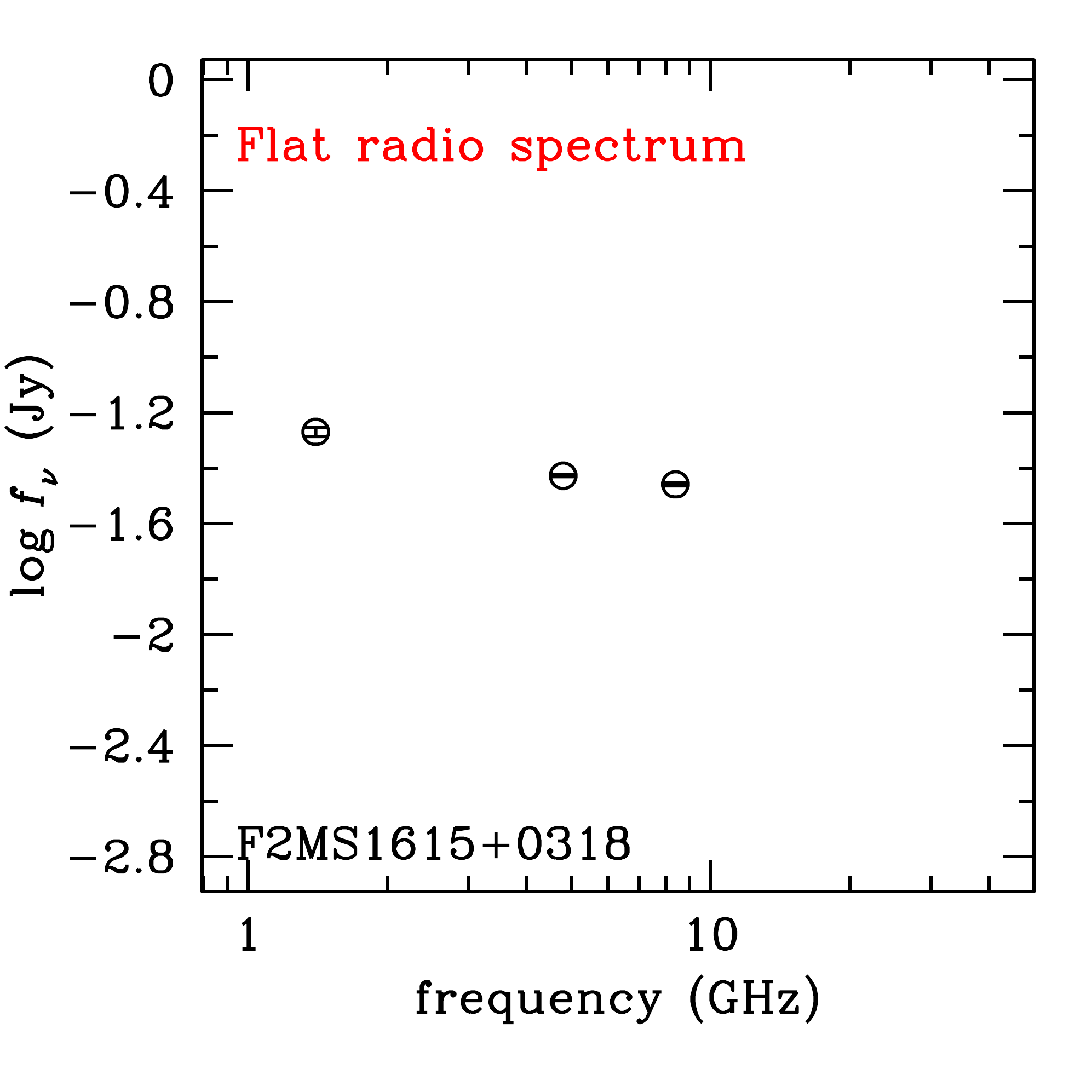}
\includegraphics[height=0.5\columnwidth]{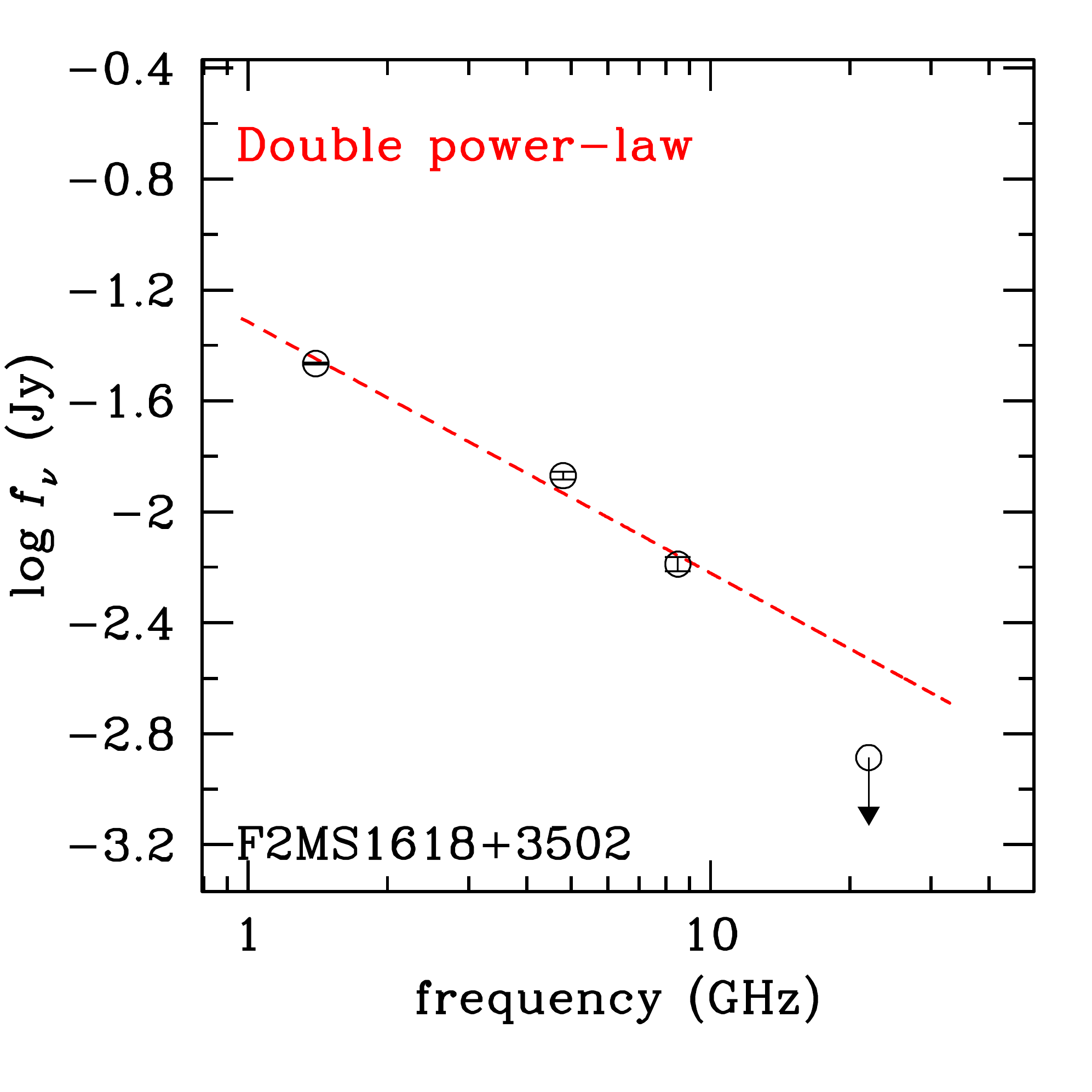}
\includegraphics[height=0.5\columnwidth]{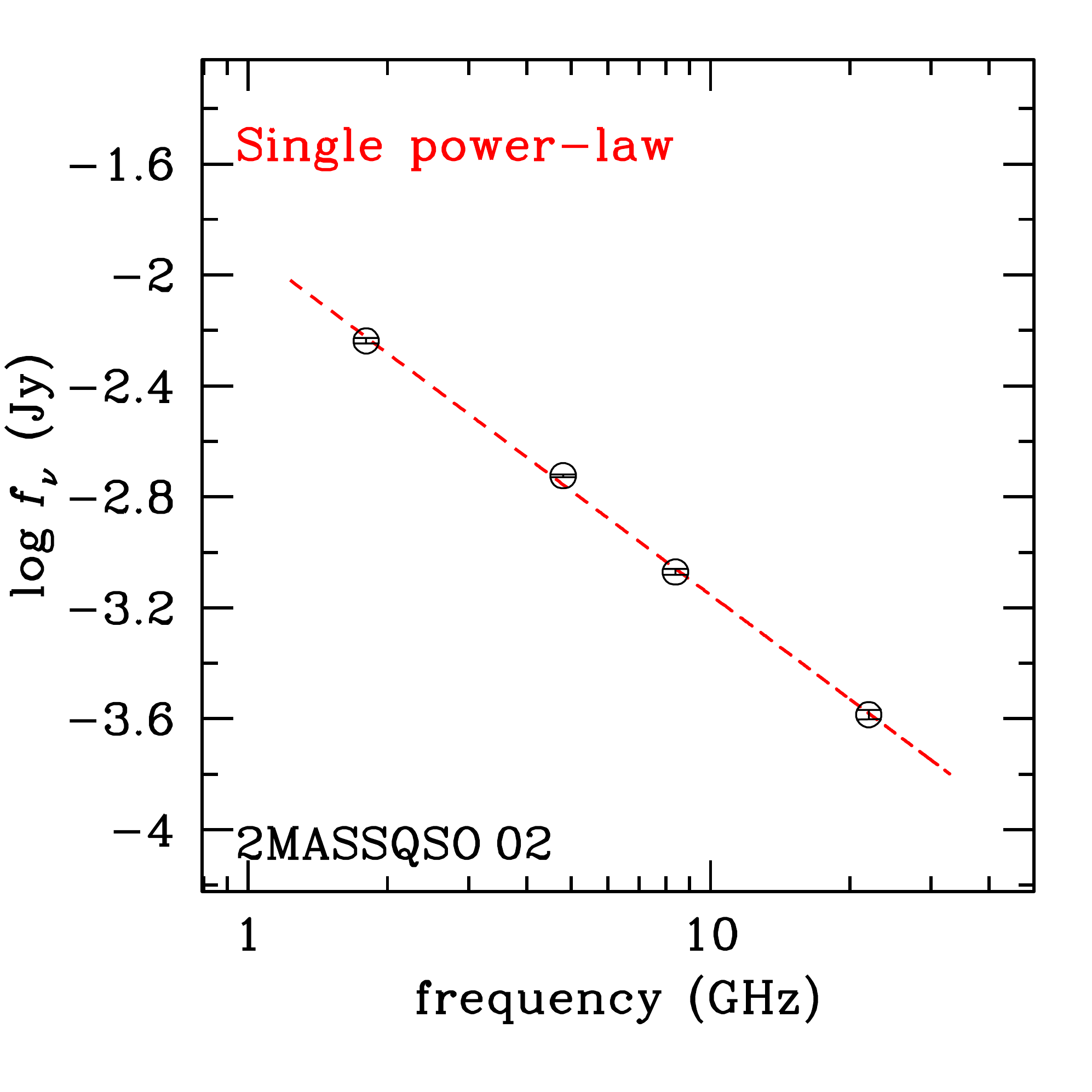}
\end{center}
\caption{Radio  spectra  of  reddened  QSOs  in  the  frequency  range
1.4-22\,GHz. Sources are  ordered in a way that  their Right Ascension
is increasing from left to right  and from top to bottom. Upper limits
in radio flux density are  shown with an arrow pointing downward.  The
y-axis in all  plots is scaled to span  the range $\pm1.5$\,dex around
the  flux  density of  each  source  at  4.8\,GHz.  The  dashed  lines
correspond to  the best-fit  single power-law or  continuous injection
model with no adiabatic loses  (Kardashev 1962). The latter is plotted
for sources  for which a single  power-law provides a poor  fit to the
data  ($P_{SPL}<0.05$, see  Table \ref{tab_slopes}).   Sources  with a
turnover at low frequencies cannot  be fit by the continuous injection
model.   Also,  Source FM2S\,1618+3502,  which  shows  evidence for  a
spectral  break in  its radio  spectrum  (see $\alpha_{\rm  1.4 -  4.8
GHz}$, $\alpha_{\rm  8.4 - 22  GHz}$ in Table  3), is not  detected at
22\,GHz.  This source  is not fit with the  continuous injection model
and only.  For  that source only the single  power-law model is shown.
In the  fit process the 22\,GHz upper  limit is not used.  he label on
the top of  each panel indicates the spectral  characteristics of each
source (single power-law, double  power-law, low frequency turnover or
flat radio spectrum).}\label{fig_spectra}
\end{figure*}

The radio Spectral Energy Distributions (SEDs) of 2MASS QSOs are shown
in  Figure  \ref{fig_spectra}.   Table \ref{tab_slopes}  presents  the
radio spectral indices $\alpha_{\rm  1.4 - 4.8 GHz}$, $\alpha_{\rm 8.4
-  22  GHz}$  between  the frequencies  1.4/4.8\,GHz  and  8.0/22\,GHz
respectively, which  provide a model independent  way of investigating
spectral features  characteristic of young jets,  e.g. spectral breaks
and/or turnovers.   Depending on the  radio spectral index at  low and
high  frequencies  and  the  overall  shape  of  the  SEDs  in  Figure
\ref{fig_spectra}  the sources  in the  sample can  be grouped  into 3
broad categories.  There are spectra  for which the radio flux density
(i)  declines monotonically  with increasing  frequency, (ii)  shows a
peak  at  GHz frequencies  and  decreases  at  both lower  and  higher
frequencies,  (iii) remains  nearly  constant in  the frequency  range
1.4--22\,GHz and cannot  be placed into one of  categories above.  The
classification scheme above is shown in Table \ref{tab_slopes}.  Class
(iii)  sources  have radio  spectra  suggestive  of free-free  thermal
emission typically found at QSO cores.

The radio SEDs of  group (i) can be described by either  a single or a
double power-law.  The data are  first fit with single power-law model
of  the  form  $S_{\nu}\propto\nu^{-\alpha}$,  where $\alpha$  is  the
spectral index.  The standard  $\chi^2$ minimisation method is adopted
for the fits. For the calculation of the $\chi^2$ we add to the formal
uncertainties  determined  by  the  JMFIT  of AIPS  listed  in  Tables
\ref{tab_sample_vla} and  \ref{tab_sample_evla} an error corresponding
to 5 per cent of the  total flux. This is to account for uncertainties
in the overall calibration of  the radio observations. The goodness of
fit  of  the  model  is  determined  by  estimating  the  probability,
$P_{SPL}$, of getting by chance  the calculated $\chi^2$ for the given
degrees  of freedom  (typically 2).   We choose  to reject  the single
power-law  model if $P_{SPL}<0.05$.   The results  are shown  in Table
\ref{tab_slopes}.

Sources for which  the single power-law model does  not provide a good
fit to the data are also  fit with a double power-law.  The continuous
injection  model  with no  adiabatic  losses \citep{Kardashev1962}  is
adopted.  In this case the spectral indices are $\alpha_{inj}$ (with a
canonical value  of about 0.7) below the  break frequency ($\nu_{br}$)
and $\alpha\approx\alpha_{inj}+0.5$ above  $\nu_{br}$.  This model was
chosen  because   it  fits  well   the  spectra  of   GPS/CCS  sources
\citep{Murgia1999,  Murgia2003}.   The  model  has 3  free  parameters
($\nu_{br}$,  $\alpha_{inj}$, normalisation) and  requires at  least 4
points in the radio SED.  Source FM2S\,1618+3502, which shows evidence
for a spectral break in its  radio SED, is not detected at 22\,GHz. We
choose  not  to  fit  this  source with  the  double  power-law  model
described above.  The best-fit parameters values, minimum $\chi^2$ and
the probabilities  of the  goodness of fit,  $P_{DPL}$, are  listed in
Table  \ref{tab_slopes}.   The  continuous injection  model  typically
provides  adequate  fits to  the  radio  spectra  of 2MASS  QSOs  with
$\alpha_{inj}\approx  0.4-0.8$ and  $\nu_{br}\approx  4-7 \rm  \,GHz$.
One exceptions  is source F2MS\,1341+3301 for which  the steepening of
the spectrum at high frequencies  is more pronounced than that assumed
by the continuous injection model ($\alpha\approx\alpha_{inj}+0.5$).

Our analysis  shows that more than  half of the sources  in the sample
(9/16) show deviations  from the single power-law SED  model and their
radio spectra can be better described by (i) a double power-law with a
steeper high  frequency index or  (ii) a turnover at  low frequencies.
These properties  are consistent  with those of  powerful ($P_{5\,GHz}
\ga \rm 10^{27}\,W/Hz$)  GPS sources (e.g.  O'Dea 1998,  Murgia et al.
1999,  2003,   Randall  et  al.   2011).   This   population  is  also
characterised by compact radio sizes, typically $\la 1$\,kpc (O'Dea et
al.   1998). The  radio data  presented in  this paper  set  only weak
limits on  the physical extent of  the radio jets.   All reddened QSOs
with  GPS  radio spectral  characteristics  appear  unresolved in  the
VLA/EVLA images. At the  resolution of those observations ($\approx 1$
and 4\,arcsec at 22\,GHz respectively) the upper limits on the size of
the  radio emitting  region are  in  the range  5-30\,kpc (see  Tables
\ref{tab_sample_vla}  and  \ref{tab_sample_evla}).   With the  present
dataset it is therefore not  possible to comment on the compactness of
the radio sources of 2MASS QSOs relative to GPS sources. 

For the reddened QSOs in the  sample that can be described by a single
power-law   (4/16)  our   analysis  yields   steep   spectral  slopes,
$\alpha\approx1$. These  sources are  also unresolved in  the VLA/EVLA
images  indicating radio  size  upper limits  $\rm  \la 30\,kpc$  (see
Tables  \ref{tab_sample_vla} and  \ref{tab_sample_evla}).   Both these
properties are similar to those  of CSS sources which show steep radio
spectra and  are typically  confined within the  extent of  their host
galaxies \citep{ODea1998, Randall2011}.

We conclude that  the majority of reddened QSOs  studied in this paper
(13/16) have  radio spectral properties  typical to those  of powerful
($P_{5\,GHz}  \ga  \rm  10^{27}\,W/Hz$)  GPS/CSS  sources,  which  are
believed to host young and  expanding radio jets.  This is interesting
because at bright fluxes  the GPS/CSS population represents only about
20-30 per  cent of  radio selected samples  (O'Dea et al.   1998).  We
caution however, that  this fraction has not been  determined for QSOs
with  radio emission at  the mJy  flux density  levels of  the present
sample.

For those 2MASS QSOs for which the continuous injection model provides
adequate   fits  to   their  radio   spectra,  the   break  frequency,
$\nu_{break}$,  listed  in  Table  \ref{tab_slopes}  can  be  used  to
estimate    their    synchrotron    ages    \citep[e.g.][]{Murgia1999,
  Drake2004_radio}

\begin{equation}\label{eq_time}
t_{syn}=0.045\,B^{-3/2}\, \bigl[\, \nu_{br} \, (1+z) \bigr],
\end{equation}

\noindent where $B$ is the magnetic field and $z$ the source redshift.
For this  calculation we will  assume the equipartition value  for the
magnetic field \citep[e.g.][]{Pacholczyk1970, Duric1991}

\begin{equation}\label{eq_magnetic}
B=\bigl[6\pi \, (1+k) \, c_{12} \, V^{-1} \, L\bigr]^{2/7},
\end{equation}

\noindent where $k$ is the heavy element to particle ratio, $V$ is the
volume, $L$ is the synchrotron luminosity between frequencies $\nu_1$,
$\nu_2$  and $c_{12}$  is  a  constant which  depends  on $\nu_1$  and
$\nu_2$  and  the  spectral  index  $\alpha_{inj}$ of  the  radio  SED
\citep[e.g.][]{Irwin1999}.  The volume of the source is estimated from
the highest  resolution data  available, which in  most cases  are the
22\,GHz observations.   In the absence of milli-arcsec  radio data for
the majority of the sources, this  is an upper limit as 2MASS QSOs are
typically unresolved in arcsec scale radio observations. We use $k=0$,
$\nu_1=100$\,MHz and $\nu_2=100$\,GHz, which  are often adopted in the
literature.   The estimated  synchrotron ages  are  somewhat sensitive
(factor of a few) to those parameters. 

 Table \ref{tab_tsyn}  lists the synchrotron ages  and luminosities as
well as  the magnetic fields for  those sources in the  sample that an
estimate  of $\nu_{br}$  can be  obtained from  the  observations.  In
addition to  sources with radio  spectra better described by  a double
power-law, we also estimate $t_{syn}$ for 2MASS QSOs F2MS1540+4923 and
F2MS1030+5806.   The former is  described by  a single  power-law with
$\alpha \approx  0.8$, similar to the canonical  value of $\approx0.7$
expected      for     a      zero     age      electron     population
\citep{Kardashev1962}. Therefore  the break frequency  for this object
likely lies above 22\,GHz at the observer's frame.  The radio spectrum
of  F2MS1030+5806 shows a  turnover below  about 5\,GHz  and therefore
cannot  be  fit with  the  continuous  injection  model. However,  the
spectral index of this source  at high frequencies is $\alpha_{\rm 8.4
- 22  GHz}=0.52\pm0.01$, close to the canonical  synchrotron slope and
similar to  the injection  slopes estimated in  Table \ref{tab_slopes}
for others 2MASS QSOs with  spectra better described by the continuous
injection model.   It is thus  likely that $\nu_{br}>22\rm  \,GHz$ for
this source.

The upper limits  of the synchrotron ages of  2MASS QSOs are typically
of  the order  of few  $10^{6}$\,yrs.  For  comparison  extended radio
galaxies    typically    have   $t_{syn}\approx    10^{7}-10^{8}$\,yrs
\citep[e.g.][]{Klein1995,   Parma1999},  while  GPS/CSS   sources  are
believed to be much younger, $t_{syn}=10^{3}-10^{5}$\,yr (e.g.  Murgia
et al. 1999).   The main limitation in the  calculation of synchrotron
ages for 2MASS QSOs is the lack of high resolution radio data. For the
one source  in the sample  (F2MS1030+5806) that such  observations are
available,  much  younger   ages  are  estimated,  $t_{syn}<2000$\,yr,
similar to  GPS/CCS sources.   It is therefore  likely that  the upper
limits in  $t_{syn}$ listed in Table  \ref{tab_tsyn} are conservative.
In  any case,  the  similarity  of the  radio  spectral properties  of
GPS/CSS sources  and 2MASS  QSOs supports the  youth scenario  for the
latter population.

\begin{figure}
\begin{center}
\includegraphics[height=0.9\columnwidth]{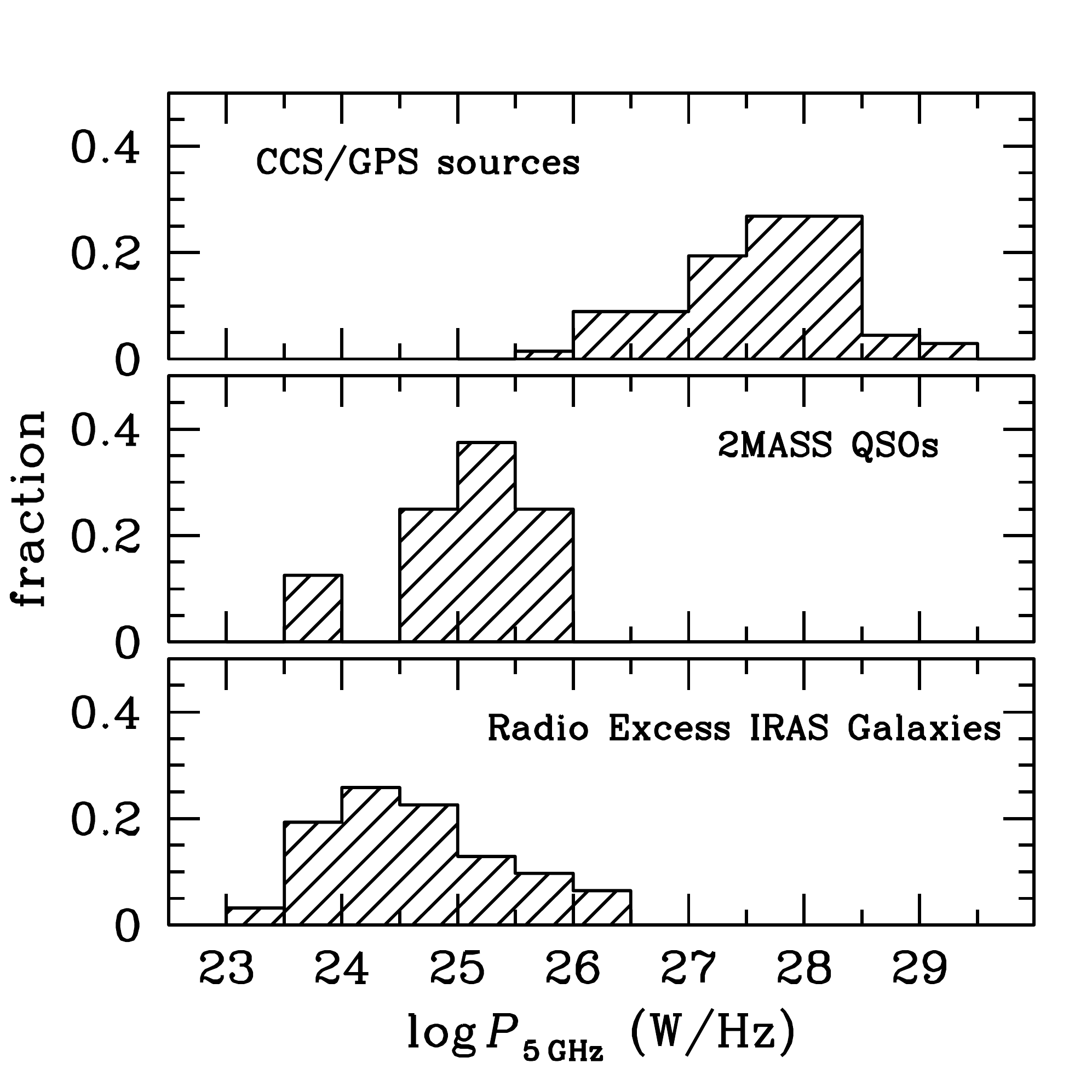}
\end{center}
\caption{Radio (5\,GHz) power  distribution for GPS/CSS sources (O'Dea
  \& Baum  1997;  top panel),  reddened QSOs  (middle panel)  and radio
  excess IRAS galaxies (Drake et  al. 2004a; bottom panel).  The radio
  powers of  GPS/CSS are  from O'Dea \& Baum (1997). For  radio excess
  IRAS galaxies the radio multifrequency radio data presented by Drake
  et al.  (2004a)  are used to interpolate and  estimate the rest-frame
  5\,GHz radio  power. The  same approach is  used for  reddened QSOs.
}\label{fig_l5}
\end{figure}

\section{Radio vs multiwavelength properties}

Next we study  the multiwavelength properties of the  reddened QSOs in
Table   \ref{tab_slopes}   in  relation   to   their  radio   spectral
characteristics. This is to investigate on a source to source basis if
the young  radio jet  scenario is supported  by observations  at other
wavelengths.

Four  sources  in Table  \ref{tab_slopes}  have  been investigated  by
Urrutia et al.   (2008) for Broad Absorption Lines  (BALs) blueward of
the $\rm MgII\,2800\AA$  emission.  Two of then are  classified as Low
Ionisation BALs (LoBALs;  F2MS\,1344+2839, F2MS\,1456+0114), the class
of BALs which are proposed to represent an early evolutionary stage in
the    quasar    lifetime    \citep[e.g.][]{Becker2000,    Trump2006}.
Interestingly both  those sources also  have radio spectra  similar to
GPS/CSS sources,  consistent with  the youth scenario.   Moreover, the
two  sources that  are characterised  as  non-BALs by  Urrutia et  al.
(2009;  F2MS\,0832+0509,  F2MS\,1012+2825)  have flat  radio  spectra,
indicative of thermal emission often observed at QSO cores.

Another source in  our sample that shows evidence  for outflows in the
form  of  double-peaked  narrow   emission  lines  separated  by  $\rm
600\,km\,s^{-1}$  \citep{Urrutia2008}   is  2MASSQSO\,10.   The  radio
spectrum  of  that  system  is  better  described  by  the  continuous
injection model, suggestive of young radio jets.

HST   observations  are  available   for  reddened   QSOs  2MASSQSO10,
F2MS0841+3604,  F2MS0915+2418  and F2MS1012+2825  \citep{Urrutia2008}.
Those sources which have radio spectral properties similar to those of
GPS/CSS sources  (2MASSQSO10, F2MS0841+3604, F2MS0915+2418)  also show
higly   disturbed   optical   morphologies   indicative   of   ongoing
mergers. The one  source that shows the least  disturbed optical light
profile is F2MS1012+2825, which is  also characterised by a flat radio
spectrum.   Although HST reveals  two nuclei  with separation  of only
0.15\,arcsec (1.2\,kpc), the optical isophotes of this object follow a
smooth elliptical light profile. 

The  reddened QSOs  in  Table 1  for  which X-ray  data are  available
\citep[][F2MS0841+3604,  F2MS0915+2418 and F2MS1012+2825]{Urrutia2005}
typically  show  hard  X-ray  spectra, indicative  of  obscured  ($\rm
N_H\ga10^{22} \,  cm^{-2}$) AGN. We  do not find any  obvious relation
between their X-ray properties and radio spectra.

Overall the optical spectra and morphologies of the reddened QSOs with
radio properties  suggestive of young  radio jets are  also consistent
with objects in  the process of formation.  Two  of the sample sources
with flat radio spectra do not share those properties.

\section{Discussion}\label{sec_discussion}

We find  that  a large  fraction  of reddened  QSOs show  either
spectral breaks  and/or turnovers at  low frequencies (9/16)  or steep
radio  spectra  and  relatively  small radio  sizes  (4/16).   Similar
spectral characteristics are observed in GPS/CSS radio sources and are
interpreted  as  evidence  for  young ($\rm  10^{3}-10^{5}\,yrs$)  and
expanding   radio    jets   \citep{ODea1997,   ODea1998,   Murgia1999,
Murgia2003, Randall2011}.  Compared  to that population however, 2MASS
QSOs are underluminous at radio  frequencies (5\,GHz, see Tables 1, 2)
by at  least few orders of  magnitude. This is  demonstrated in Figure
\ref{fig_l5} showing  the distribution of 5\,GHz radio  powers for our
sample and GPS/CSS sources \citep{ODea1997}. This is likely because of
differences in  both the  flux density limits  ($S_{1.4}\ga$5\,mJy for
2MASS  QSOs  vs  $S_{4}\ga$1\,Jy  for  GPS/CSS sources)  and  the  the
redshift distributions (median redshift of  0.74 for 2MASS QSOs vs 0.9
for GPS/CSS sources) of the two  samples.  In any case, the 2MASS QSOs
presented in  this paper appear to  be low power  analogues of GPS/CCS
sources. 

The comparison between the radio  properties of 2MASS QSOs and GPS/CSS
sources should also  take into account the size  of the radio emitting
region in the two populations.   Although CCS sources extend over tens
of kpc, the GPS radio  population is more compact ($\la 1$\,kpc; O'Dea
et al.  1998). The VLA/EVLA  observations presented in this paper lack
the spatial  resolution to set interesting  limits on the  size of the
radio emitting region of 2MASS  QSOs.  Nevertheless, for the one 2MASS
QSOs in our  sample (F2MS1030+5806) that tight constraints  on the jet
size are available from milli-arcsec resolution observations, both its
extent  and  infered  radio  age  are consistent  with  those  of  GPS
sources.  It is  interesting that  this sources  also has  the highest
radio power in  the sample, $P_{\rm 5\,GHz}=10^{26} \rm  \, W/Hz$ (see
Table \ref{tab_sample_vla}).

The high  incidence of radio  spectral properties suggestive  of young
radio  jets among reddened  QSOs is  consistent with  suggestions that
these  systems represent an  early and  brief stage  of the  growth of
SMBHs      \citep{Hopkins2008_sam,      Urrutia2008,      Urrutia2009,
Georgakakis2009}. In  this picture reddened QSOs  are the predecessors
of UV/optically  bright type-I  AGN. They are  captured at  a critical
time of  their evolution  during which the  natal dust and  gas cocoon
that enshrouds the SMBH is blown away by some feedback process related
to the energy output of the central engine.

Outflows in the ionised gas are indeed observed in reddened QSOs.  For
the high redshift  ($z\ga0.9$) subset of those sources,  for which the
Mg\,II\,2800\AA\, emission line  lies within the optical spectroscopic
window,  a  high  fraction  of  low ionisation  BALs  is  reported  by
\cite{Urrutia2009}.  The  young radio jets  found by our  analysis may
play an  important role in driving  these outflows. In  the more radio
luminous  GPS/CSS sources,  outflows  also traced  in  the ionised  or
neutral  gas \citep[e.g.][]  {Holt2003, Holt2006,  Holt2008, Holt2011,
  Morganti2005} and  are proposed to be associated  with the expanding
powerful  radio jets  \citep{Batcheldor2007}.  These  findings coupled
with the suggestion by Georgakakis  et al.  (2009) for a high fraction
of  radio  detections among  reddened  QSOs  compared to  UV/optically
bright ones, highlights the potential  significance of radio jets as a
feedback mechanism at the early stages of AGN evolution.

This  is consistent  with the  two-stage feedback  scheme  proposed by
\cite{Hopkins_Elvis2010}.  They suggest that some AGN related process,
which the evidence above suggests might be radio jets, first drives
a wind in the warm/hot diffuse gas.  Cold clouds are then deformed and
shredded  in the wake  of this  outflow, thereby  effectively becoming
more  susceptible to pressure  and ionisation  from the  QSO intense
radiation  field.  The  net result  of this  process is  a significant
reduction of the  initial AGN energy output that  needs to couple with
the ISM to  blow away or destroy  the cold gas of its  host galaxy, in
better  agreement  with  recent observations  \citep[e.g.][]{Holt2006,
  Holt2011}.

Another population of  sources which are also found  to include a high
fraction of relatively young radio jets ($\rm <10^{6}\,yrs$) are Radio
Excess  IRAS Galaxies  \citep{Drake2003, Drake2004_radio}.   Given the
similarities in the radio properties of Radio Excess IRAS Galaxies and
reddened 2MASS  QSOs it is interesting to  compare the multiwavelength
properties of the two populations.

Radio Excess IRAS  Galaxies are selected to have  radio emission above
the   expectation   from   the  radio/far-infrared   correlation   for
star-formation \citep{Drake2003}.   The majority of  these sources (70
per cent) are  hosted by galaxies which either  show disturbed optical
morphology or have nearby  companions suggestive of tidal interactions
\citep{Drake2004_hosts}.   This  is similar  to  2MASS  QSOs, a  large
fraction   of   which   show   morphological  evidence   for   mergers
\citep{Urrutia2008}. Studies  of the  optical spectra of  Radio Excess
IRAS   Galaxies  \citep{Buchanan2006}   show  a   large   fraction  of
post-starburst   stellar  continua,  indicating   recently  terminated
($<1$\,Gyr)  star-formation activity.  In  reddened QSOs,  the optical
spectra have  a significant contribution from the  central engine, and
therefore  the  identification of  host  galaxy  spectral features  is
difficult.   Nevertheless,  Georgakakis  et  al.  (2009)  studied  the
infrared  properties of  reddened 2MASS  QSOs and  found  evidence for
enhanced   star-formation  compared   to   optically  selected   QSOs.
Blueshifts between the [O\,III]\,5007\AA\, line and the broad emission
lines  are  also  observed  in  Radio Excess  IRAS  Galaxies  and  are
interpreted as evidence for interactions between the radio jet and the
surrounding  interstellar   medium  \citep{Buchanan2006}.   Similarly,
outflows   in   the  warm   ISM   are   observed   in  reddened   QSOs
\citep{Urrutia2009}.    Figure  \ref{fig_l5}   shows   that  the   two
populations  have similar  radio power  distributions at  5\,GHz, with
Radio  Excess  IRAS Galaxies  having  a  lower  median.  This  can  be
attributed to the different redshift distributions of the two samples,
with medians of 0.15 and 0.74 for Radio Excess IRAS Galaxies and 2MASS
QSOs, respectively.

\begin{table*}
\caption{The sample of reddened QSOs observed by VLA}\label{tab_sample_vla}
\footnotesize
\begin{center}
\begin{tabular}{l cc cc ccc cc}
\hline
ID &
$\alpha$ &
$\delta$ &
redshift &
$S_{\rm 1.4\,GHz}$ & 
$S_{\rm 4.8\,GHz}$ & 
$S_{\rm 8.4\,GHz}$ & 
$S_{\rm 22.0\,GHz}$ & 
$\log P_{\rm 5\,GHz}$ &
beam size\\
 &
(J2000)&
(J2000)&
     &
(mJy)&
(mJy)&
(mJy)&
(mJy)& 
(W/Hz)&
(kpc)\\

(1)&
(2)&
(3)&
(4)&
(5)&
(6)&
(7)&
(8)& 
(9)&
(10)\\

\hline
 2MASSQSO10    & 08:25:02.05 &+47:16:51.96& 0.804 & $49.2\pm2.2$ & $26.8\pm0.6$  &  $16.2\pm0.8$ &  $4.8\pm0.7$ & 25.4 & 7.5\\

 F2MS0832+0509 & 08:32:11.64 &+05:09:01.04& 1.070 & $31.3\pm1.2$ &
 $17.2\pm0.5$  &  $19.5\pm0.5$ &  $19.5\pm0.8$ & 25.8 & 8.1\\

 F2MS0932+3854 & 09:32:33.29 &+38:54:28.12& 0.506 & $51\pm2$     &
 $30.0\pm0.6$  &  $21.0\pm0.5$ &  $8.8\pm 1.7$ & 25.2 & 6.1\\


 F2MS1030+5806 & 10:30:39.62 &+58:06:11.41& 0.504 & $89.0\pm9.5$ &
 $132.0\pm0.5$ &  $142.0\pm0.5$&  $84.32\pm0.72$ & 26.0 & 6.1\\ 

 F2MS1248+0531 & 12:48:47.16 &+05:31:30.79& 0.740 & $16.6\pm1.2$ &
 $4.9\pm0.5$   &  $3.08\pm0.43$&  $<1.52$ & 24.6 & 21.9  \\

 F2MS1341+3301 & 13:41:08.11 &+33:01:10.23& 1.720 & $65.1\pm0.6$ &
 $36.6\pm0.4$  &  $21.9\pm0.5$ &  $6.2\pm0.9$ & 25.9 & 8.5\\ 


 F2MS1540+4923 & 15:40:43.74 &+49:23:23.89& 0.696 & $32.99\pm0.0$&
 $12.50\pm0.40$& $8.6\pm0.5$   &  $4.4\pm0.6$ & 25.0 & 7.1 \\ 

 F2MS1615+0318 & 16:15:47.90 &+03:18:50.83& 0.424 & $54\pm2$     &
 $37.4\pm0.4$  & $34.8\pm0.5$  &     --  & 25.2   & 16.8  \\

 F2MS1618+3502 & 16:18:09.72 &+35:02:08.53& 0.446 & $34.3\pm0.2$ &
 $13.5\pm0.4$  & $6.5\pm0.4$   & $<1.30$ & 24.6 & 17.1\\
\hline
\end{tabular}

\begin{list}{}{}
\item The columns  are: (1) Source ID as listed  in Georgakakis et al.
(2009) or Urrutia et al. (2009); (2) right ascension; (3) declination;
(4) source redshift; (5) 1.4\,GHz  radio flux density in mJy.  Sources
F2MS1540+4923 and  F2MS1618+3502 have  not been observed  at 1.4\,GHz.
The listed  flux densities are  from the FIRST  survey (F2MS1540+4923)
and  Glikman et al.   (2007; F2MS1618+3502);  (6) 4.8\,GHz  radio flux
density in mJy; (7) 8.4\,GHz  radio flux density in mJy; (8) 22.0\,GHz
radio flux density in mJy.  Source F2MS1615+0318 has not been observed
at this  frequency because of  technical problems; (9) Radio  power at
5\,GHz estimated  by interpolating  between the observed  data points;
(10) Linear size in kpc of the antenna beam at 22\,GHz at the redshift
of the source. For sources not detected or not observed at 22\,GHz the
8\,GHz  beam  size  is   listed.   For  unresolved  sources  (all  but
F2MS1618+3502, see section 2.1) this  the upper limit of the extent of
the radio emitting region.
\end{list}

\end{center}
\end{table*}

\begin{table*}
\caption{The sample of reddened QSOs observed by EVLA}\label{tab_sample_evla}
\footnotesize
\begin{center}
\begin{tabular}{l ccc cc ccc cc}
\hline
ID &
$\alpha$ &
$\delta$ &
redshift &
$S_{\rm 1.3\,GHz}$ & 
$S_{\rm 1.8\,GHz}$ & 
$S_{\rm 4.8\,GHz}$ & 
$S_{\rm 8.4\,GHz}$ & 
$S_{\rm 22.0\,GHz}$& 
$\log P_{\rm 5\,GHz}$ &
beam size\\
 &
(J2000)&
(J2000)&
     &
(mJy)&
(mJy)&
(mJy)&
(mJy)&
(mJy)&
(W/Hz)&
(kpc)\\

(1)&
(2)&
(3)&
(4)&
(5)&
(6)&
(7)&
(8)& 
(9)&
(10)\\
\hline

F2MS0841+3604  & 08:41:04.98 &+36:04:50.09& 0.552 & $6.6\pm0.2^{}$& -
& $9.5\pm0.2$   &  -            & $1.08\pm0.02$ & 24.6 & 25.0\\ 

F2MS0915+2418  & 09:15:01.71 &+24:18:12.24& 0.842 & $10.1\pm0.1^{}$& -
& $21.96\pm0.07$& $13.97\pm0.09$& $5.34\pm0.07$ & 25.4& 30.5\\ 

F2MS1012+2825  & 10:12:30.49 &+28:25:27.15& 0.937 & $8.5\pm0.1$   &
$8.3\pm0.2$   & $8.3\pm0.1$   & $5.64\pm0.04$ & $3.64\pm0.03$& 25.1 & 31.5\\ 

2MASSQSO06     & 13:40:39.68 &+05:14:19.90& 0.259 & $12.9\pm0.3$&
$13.4\pm0.3$             & $4.46\pm0.03$ & $2.89\pm0.03$ &
$1.02\pm0.01$ & 23.8 & 16.0\\ 

F2MS1344+2839  & 13:44:08.31 &+28:39:31.97& 1.770 & $10.5\pm0.2$   & -
& $3.99\pm0.07$ & $2.11\pm0.04$ & $0.68\pm0.01$ & 25.0 & 33.8\\ 

F2MS1456+0114  & 14:56:03.09 &+01:14:45.71& 2.378 & $8.7\pm0.2^{}$& -
& $11.68\pm0.06$& $8.21\pm0.04$ & $2.64\pm0.02$ & 25.7 & 32.6 \\ 

2MASSQSO02     & 20:56:29.76 &-06:50:55.40& 0.635 & -              &
$5.8\pm0.1$ & $1.89\pm0.02$ & $0.85\pm0.02$ & $0.26\pm0.01$ & 24.0 & 27.4\\ 

\hline  
\end{tabular}

\begin{list}{}{}
\item The columns  are: (1) Source ID as listed  in Georgakakis et al.
(2009) or Urrutia et al.  (2009; (2) right ascension; (3) declination;
(4) source redshift;  (5) 1.3\,GHz radio flux density  in mJy.  Source
F2MS1456+0114 has a  nearby radio source which is  not resolved by the
EVLA L-band observations.  The radio flux density at this frequency is
from   the   FIRST.   The   L-band   observations  of   F2MS0841+3604,
F2MS0915+2418 suffered technical  problems.  The listed flux densities
for those  objects are from the  FIRST.  The 1.3\,GHz  flux density of
2MASSQSO02 could not be estimated;  (6) 1.8\,GHz radio flux density in
mJy.   The  1.8\,GHz  flux  density  of  F2MS1344+2839  could  not  be
estimated; (7) 4.8\,GHz radio flux  density in mJy; (8) 8.4\,GHz radio
flux  density in mJy;  (9) 22.0\,GHz  radio flux  density in  mJy; (9)
Radio power at 5\,GHz  estimated by interpolating between the observed
data points; (10) Linear size in kpc of the antenna beam at 22\,GHz at
the redshift of the source.  This  is the upper limit of the extent of
the radio emitting region as  all listed sources are unresolved in the
EVLA images.
\end{list}

\end{center}
\end{table*}

\begin{table*}
\caption{Radio spectral slopes}\label{tab_slopes}
\footnotesize
\begin{center}
\begin{tabular}{l cc cc cc c c  c c c}
\hline
ID & 
$\alpha_{\rm 1.4\, GHz}^{\rm 4.8\,GHz}$ & 
$\alpha_{\rm 8.5\, GHz}^{\rm 22\,GHz}$ & 
class &
\multicolumn{3}{c}{SPL Parameters} & \multicolumn{4}{c}{DPL Parameters}&  
GPS/CCS \\

  & 
  & 
  &
  &
 $\alpha$ & $\chi^2$ & $P_{SPL}$  & $\nu_{break}$ & $\alpha_{inj}$ & $\chi^2$    & $P_{DPL}$ & \\  

  & 
  & 
  &
  &
  & &   & (GHz) &  &     & & \\

\hline

2MASSQSO10    & $0.49\pm0.07$ & $1.3\pm0.2$   & (i)   & $0.75_{-0.02}^{+0.03}$ & 27.0 & $<10^{-4}$ & $5.1_{-1.1}^{+1.1}$ & $0.52_{-0.07}^{+0.06}$&  2.5 & 0.11 & DPL\\

F2MS0832+0509 & $0.49\pm0.07$ & $0.00\pm0.09$ & (iii) &      --              & --   &   --       & --                  &      --               & --   & --    & \\

F2MS0841+3604 & $-0.30\pm0.06$&    --         &  (ii) &      --              &  --  &   --       & --                  &      --               & --   &  --   & LFT \\

F2MS0915+2418 & $-0.63\pm0.06$& $1.00\pm0.07$ & (ii)  &      --              &  --  &  --        & --                  &      --               & --   &  --   & LFT\\

F2MS0932+3854 & $0.43\pm0.07$ &  $0.9\pm0.2$  & (i)   & $0.54_{-0.04}^{+0.04}$ & 6.7  &  $0.035$   & $6.7_{-0.9}^{+0.9}$ & $0.43_{-0.05}^{+0.05}$&  0.1 & 0.75 & DPL\\ 

F2MS1012+2825 & $0.02\pm0.06$ & $0.45\pm0.07$ & (iii)  &      --              &  --  & --         & --                  & --                    & --  & --   & \\

F2MS1030+5806 & $-0.32\pm0.10$& $0.54\pm0.07$ & (ii)  &      --              &  --  &  --        &  --                 &  --                   & --  & -- &  LFT\\
 
 F2MS1248+0531& $1.0\pm0.1$   & $>0.73$       & (i) &$0.96_{-0.08}^{+0.07}$& 0.3  & $0.58$     &   --                &  --                   &  -- &  --  & SPL\\

2MASSQSO06    & $0.88\pm0.06$ & $1.08\pm0.07$ & (i)   &$0.92_{-0.04}^{+0.05}$&  5.3 & $0.07$     &    --                &  --                   &  -- &  -- & SPL\\

F2MS1341+3301 & $0.47\pm0.06$ & $1.3\pm0.2$   & (i)   & $0.71_{-0.03}^{+0.03}$ & 37.8 & $<10^{-4}$& $5.1_{-0.7}^{+0.5}$ & $0.47_{-0.04}^{+0.02}$ &  5.0 & $0.025$  &DPL \\

F2MS1344+2839 & $0.79\pm0.06$ & $1.18\pm0.08$ & (i)   & $0.97_{-0.02}^{+0.03}$&  20.0 & $<10^{-4}$&  $3.8_{-1.2}^{+3.2}$ & $0.66_{-0.08}^{+0.08}$ &   0.04   & 0.84  & DPL\\

F2MS1456+0114  & $-0.24\pm0.06$ & $1.18\pm0.07$ & (ii)&      --               &  --   &  --       & --                     &  --                   &   --     & --  & LFT\\

 F2MS1540+4923 &  $0.78\pm0.06$ & $0.7\pm0.2$   &(i)& $0.77_{-0.02}^{+0.02}$& 0.6   & 0.74      &  --                    &  --                   &  --      & -- & SPL\\ 

F2MS1615+0318  &  $0.29\pm0.06$ &  --           & (iii)  &  --                   & --    &  --       &  --                    &  --                   & --       & -- & \\

F2MS1618+3502  & $0.76\pm0.06$  & $>1.7$        & (i) & $0.88^{+0.04}_{-0.04}$& 2.21   & 0.14     &  --                   &   --                   & --       & --   & DPL\\

2MASSQSO02     & $0.91\pm0.06$  & $1.23\pm0.09$ & (i) & $1.25_{-0.04}^{+0.03}$ &  2.52 & 0.28      &  --                   &   --                   &  --     &  -- & SPL\\

\hline
\end{tabular}

\begin{list}{}{}
\item The columns are: (1): Source ID; (2) spectral index $\alpha_{\rm
1.4 - 4.8 GHz}$ between the frequencies 1.4 and 4.8\,GHz; (3) spectral
index  $\alpha_{\rm 8.4  - 22  GHz}$ between  the frequencies  8.0 and
22\,GHz respectively; (4) classification based on the spectral indices
$\alpha_{\rm 1.4\, GHz}^{\rm  4.8\,GHz}$, $\alpha_{\rm 8.5\, GHz}^{\rm
22\,GHz}$ and  the overall shape of  the radio spectrum  (see text for
details); (5): best-fit spectral index for the single power-law model.
The uncertainties correspond to the  68 per cent confidence level. For
source F2MS1618+3502 only the 1.4--8.4\,GHz  data are used in the fit;
(6) $\chi^2$  of the  best-fit single power-law  model and  degrees of
freedom;  (7) goodness  of fit  probability for  the  single power-law
model  for   the  estimated  minimum  $\chi^2$  and   the  degrees  of
freedom. For  sources with $P_{SPL}<0.05$ best-fit  parameters for the
continuous injection  model (double power-law) are  also estimated. In
this  case the additional  columns are  (8) best-fit  and 68  per cent
confidence  level errors  for the  break frequency  of  the continuous
injection model; (9) best-fit and  68 per cent confidence level errors
for  the low  frequency  spectral index  of  the continuous  injection
model; (10)  $\chi^2$ of the  best-fit continuous injection  model and
number of  degrees of  freedom; (11) probability  of goodness  of fit;
(11) marks sources with  radio spectral features consistent with those
of the  GPS/CSS population. The  acronyms indicate LFT:  low frequency
turnovers, DPL: double power-law and SPL: steep power-law spectrum.
\end{list}

\end{center}
\end{table*}

\begin{table*}
\caption{Synchrotron timescales}\label{tab_tsyn}
\footnotesize
\begin{center}
\begin{tabular}{l cc cc c}
\hline
ID & source size & $\nu_{break}$  & $L_{syn}$ & $B_{min}$      & $t_{syn}$ \\ 
   & (kpc)       &   (GHz)        &  (erg/s)  &  ($\rm \mu G$)  &   (yr)   \\ 
\hline
2MASSQSO10       & $<7.5$ & $5.1$  & $4.0\times10^{43}$ & $>34.2$ & $<2.3\times10^{6}$ \\ 
F2MS0932+3854 & $<6.1$ & $6.7$  & $1.5\times10^{43}$ & $>28.4$ & $<3.0\times10^{6}$ \\
F2MS1030+5806 & 0.06   & $>22.0$& $7.1\times10^{43}$ & $2500$  & $<2.0\times10^{3}$ \\
F2MS1341+3301 & $<8.5$ & $5.1$  & $7.6\times10^{44}$ & $>69.1$ & $<0.6\times10^{6}$ \\
F2MS1344+2839 & $<33.8$ & $3.8$ & $7.0\times10^{43}$ & $>41.0$  & $<1.7\times10^{6}$ \\ 
F2MS1540+4923 & $<7.1$ & $>22.0$& $8.3\times10^{42}$ & $>12.5$ & $<9.9\times10^{6}$  \\
\hline
\end{tabular}

\begin{list}{}{}
\item The columns are: (1): Source ID; (2): source size in the highest
  resolution  radio data  available.  Those  are typically  the 22\,GHz
  observations, except from source F2MS1030+5806 for which milli-arcsec
  scale radio  data are available  from Helmboldt et al.   (2007); (3)
  best-fit  break frequency  of the  continuous injection  model; (4):
  synchrotron luminosity in erg/s between frequencies $\nu_1=100$\,MHz
  and  $\nu_2=100$\,GHz; (5):  minimum energy  (equipartition) magnetic
  field in $\rm \mu  G$ (equation \ref{eq_magnetic}); (6): synchrotron
  timescale in years (equation \ref{eq_time})
\end{list}

\end{center}
\end{table*}

\section{Conclusions}

In  summary  the multifrequency  radio  observations (1.4-22\,GHz)  of
reddened  QSOs  presented in  this  paper  reveal  a high  fraction  of
spectral features (breaks and/or  turnovers) suggestive of young ($\la
10^{6}$\,yrs) radio jets.  This  is consistent with the youth scenario
for reddened QSOs  according to which they are  systems captured at an
early stage of  the formation of their SMBHs.   The evidence for young
and expanding radio lobes coupled with claims for a higher fraction of
radio  detections  among  reddened  QSOs (Georgakakis  et  al.  2009),
suggest  that feedback  associated  with  radio jets  may  be play  an
important role at  the early phases of the evolution  of AGN and their
host galaxies.  This supports the two stage  feedback scheme suggested
by Hopkins \& Elvis (2010). 

\section{Acknowledgements}

The  authors  would  like  to  thank the  referee,  Joanna  Holt,  for
constructive  comments.  AG  acknowledges financial  support  from the
Marie-Curie  Reintegration  Grant  PERG03-GA-2008-230644.  JA  and  MG
gratefully  acknowledge  support   from  the  Science  and  Technology
Foundation    (FCT,    Portugal)    through   the    research    grant
PTDC/CTE-AST/105287/2008.  The National Radio Astronomy Observatory is
a  facility   of  the  National  Science   Foundation  operated  under
cooperative agreement by Associated Universities, Inc.

\bibliography{mybib}{}
\bibliographystyle{mn2e}

\end{document}